\newif\ifmypaper\mypapertrue
\newif\ifnotmypaper\notmypaperfalse
\ifmypaper
  \notmypaperfalse
  \documentclass[journal=aps, apsjournal=prc, myfonts=true, allwarnings=false,longbibliography]{my_paper}

  \usepackage[%matchfonts=true,  % Requires Minion Pro and Myriad Pro and lualatex.
              midcaps=true,
              euler=false,  % Use mathpazo for math
              nag=true,     % Turn warnings into errors (no footnotes, etc.)
              todo=false,
              secnoindent=true,
              ]{my_paper}
\else
  \notmypapertrue
  \documentclass[aps,prc,
                 preprint,
                 twocolumn,
                 amsmath, amssymb,
                 superscriptaddress, nofootinbib, longbibliography
                 ]{revtex4-2} 
\fi

\usepackage[T1]{fontenc}
\usepackage[utf8]{inputenc}

\usepackage[percent]{overpic}  % Allow overlaying text on images (i.e. to add "a)", b) etc.)
\usepackage{graphicx}          % For including images

\usepackage{hyperref}
\hypersetup{colorlinks=true,
            pdfpagemode=UseNone,
            linkcolor=black,
            citecolor=blue,
            urlcolor=blue}

\usepackage{xcolor} % for custom colors

\newcommand{\noacronym}[2]{\newacronym{#1}{#2}{<undefined>}\glsunset{#1}}
\newacronym{NSF}{\mysc{nsf}}{National Science Foundation}
\newacronym{HPC}{\mysc{hpc}}{high-performance computing}
\newacronym{GPU}{\mysc{gpu}}{graphics processing unit}
\newacronym{CPU}{\mysc{cpu}}{central processing unit}
\glsunset{CPU}
\noacronym{LUMI}{\mysc{lumi}}
\newacronym{SLDA}{\mysc{slda}}{superfluid local density approximation}
\newacronym{ASLDA}{\mysc{aslda}}{asymmetric superfluid local density approximation}
\newacronym{WSLDA}{\mysc{w-slda}}{Warsaw superfluid local density approximation}
\glsunset{WSLDA}
\newacronym{UFG}{\mysc{ufg}}{unitary Fermi gas}
\newacronym{TDSLDA}{\mysc{tdslda}}{time-dependent superfluid local density approximation}
\newacronym{HF}{\mysc{hf}}{Hartree–Fock}
\newacronym{HFB}{\mysc{hfb}}{Hartree–Fock–Bogoliubov}
\newacronym{DFT}{\mysc{dft}}{density functional theory}
\newacronym{TDDFT}{\mysc{tddft}}{time-dependent density functional theory}
\newacronym{TDHFB}{\mysc{tdhfb}}{time-dependent Hartree–Fock–Bogoliubov}
\newacronym{GPE}{\mysc{gpe}}{Gross-Pitaevskii equation}
\newacronym{SGPE}{\mysc{sgpe}}{stochastic Gross-Pitaevskii equation}
\newacronym{BEC}{\mysc{bec}}{Bose-Einstein condensate}
\newacronym{BCS}{\mysc{bcs}}{Bardeen Cooper Schrieffer}
\newacronym{FFT}{\mysc{fft}}{fast Fourier transform}
\newacronym{LQF}{\mysc{lqf}}{local--quantum-friction}
\newacronym{LQC}{\mysc{lqc}}{local--quantum-cooling}
\newacronym{LO}{\mysc{lo}}{Larkin-Ovchinnikov}
\newacronym{FF}{\mysc{ff}}{Fulde-Ferrell}
\newacronym{LOFF}{\mysc{loff}}{Larkin-Ovchinnikov -- Fulde-Ferrell}

\ifnotmypaper
  \boldsymbol is recommended by the APS
  \newcommand{\vect}[1]{\boldsymbol{#1}}
  \newcommand{\mat}[1]{\boldsymbol{#1}}
  \newcommand{\I}{\mathrm{i}}    % For imaginary numbers
  \DeclareMathOperator{\Tr}{Tr}  % Has better spacing than $\text{Tr}$ in some contexts
\fi

\renewcommand{\mat}[1]{\underline{\boldsymbol{#1}}}

\newcommand{\ua}{\uparrow}
\newcommand{\da}{\downarrow}

\newcommand{\eF}{\varepsilon_{F}}

\newcommand{\kF}{k_{\textrm{F}}}
\newcommand{\tot}{\mathrm{tot}}

\newcommand{\HH}{\respch{\mathscr{H}}}
\newcommand{\QQ}{\respch{\mathscr{Q}}}
\newcommand{\NN}{\respch{\mathscr{N}}}
\newcommand{\RR}{\respch{\mathscr{R}}}
\newcommand{\UU}{\respch{\mathscr{U}}}

\newcommand{\D}{\mat{\Delta}}
\newcommand{\h}{\mat{h}}
\newcommand{\A}{\mat{A}}
\newcommand{\B}{\mat{B}}

\newcommand{\td}{\mathrm{TD}}
\newcommand{\HHtd}{\respch{\mathscr{H}_\td}}
\newcommand{\htd}{\mat{h}_\td}
\newcommand{\Atd}{\mat{A}_{\td}}
\newcommand{\Btd}{\mat{B}_{\td}}

\newcommand{\Rdens}{\respch{\mathscr{R}}}

\newcommand{\diss}{^{\textrm{(d)}}}
\newcommand{\Udiss}{U\diss}
\newcommand{\Ddiss}{\Delta\diss}

\newcommand{\Nreq}{N_{\textrm{req.}}}

\let\Re\relax
\let\Im\relax
\DeclareMathOperator{\Re}{Re}
\DeclareMathOperator{\Im}{Im}

\newcommand{\black}{\color{black}}
\newcommand{\respch}[1]{{\protect\black #1}}

\begin{document}
% \title{Implementations on local cooling dynamics as a cooling mechanism}
%\title{Local Quantum Cooling with Pairing: Unitary Dissipation in Large Fermi Systems}
 \title{\respch{ Local Quantum Cooling for Large Fermi Systems with Pairing }}

\preprint{INT-PUB-25-032}
\preprint{NT@UW-26-6}

\author{J. E. Alba-Arroyo}
\email{jose.arroyo@pw.edu.pl}
\affiliation{%
  Faculty of Physics, Warsaw University of Technology,
  Ulica Koszykowa 75, 00-662 Warsaw, Poland
}

\author{Daniel P{\k e}cak}
\email{daniel.pecak@ifpan.edu.pl}
\affiliation{%
  Institute of Physics, Polish Academy of Sciences,
  Aleja Lotnikow 32/46, PL-02668 Warsaw, Poland
}

\author{Michael M\textsuperscript{c}Neil Forbes}
\email{m.forbes@wsu.edu}
\affiliation{%
  Department of Physics and Astronomy, Washington State University,
  Pullman, WA 99164, USA
}
\affiliation{%
  Department of Physics, University of Washington,
  Seattle, WA 98195-1560, USA
}

\author{Gabriel Wlaz\l{}owski}
\email{gabriel.wlazlowski@pw.edu.pl}
\affiliation{%
  Faculty of Physics, Warsaw University of Technology,
  Ulica Koszykowa 75, 00-662 Warsaw, Poland
}
\affiliation{Department of Physics, University of Washington,
  Seattle, WA 98195-1560, USA
}

\date{\today}
\begin{abstract}
  We present a framework for local quantum cooling that can be efficiently applied to large-scale Fermi systems.
  The method introduces local Hermitian operators as a cooling potential while strictly preserving the unitarity of time evolution.
  Our formulation scales favorably with system size and can be seamlessly integrated into time-dependent density functional theory frameworks.
  We demonstrate that energy cooling arises from the damping of particle currents and pairing-field fluctuations.
  Furthermore, we develop a variant of the scheme that allows the particle number to vary in time, enabling controlled density scans.
  The method is generic and versatile, as illustrated by applications to spin-imbalanced unitary Fermi gases and to nuclear matter in the neutron-star crust. The framework can be naturally extended to include stochastic noise, providing a foundation for studying 
  thermalization in strongly interacting Fermi superfluids.
\end{abstract}

\maketitle

\section{ Introduction.}

\respch{
Bridging theoretical predictions with experimentally realized strongly interacting Fermi systems requires a consistent treatment of finite-temperature effects, since such systems are inherently realized at nonzero temperatures. At finite temperatures, the presence of thermal excitations requires going beyond purely deterministic mean-field descriptions. While the theory of open quantum systems~\cite{FeynmanVernon, Caldeira, Haake:1973} provides an exact framework for incorporating dissipation, its implementation in strongly interacting many-body systems is complicated, with non-local, time-dependent behavior in which the system's dynamics are entangled with those of the reservoir. In practice, these methods are mostly limited to small systems.
}

\respch{An alternative is to employ effective descriptions in which the coupling to the environment is modeled through stochastic and damping terms. In this context, theoretical frameworks such as the stochastic Gross-Pitaevskii equation (\mysc{sgpe})~\cite{Gardiner2002, GardinerDavis2003, Stoof1999,Cockburn2009} were developed as a manageable extension of mean-field theory, successfully capturing the relaxation and thermalization dynamics of Bose systems. Although originally formulated for a single macroscopic wavefunction occupying the lowest-energy mode, the \mysc{sgpe} provides conceptual guidance for generalizations to fermionic systems, where one typically works with an orthogonal set of single-particle wavefunctions.}

\respch{In the fermionic case, the analogue of the \mysc{sgpe} can be expressed as an augmented Schr\"odinger equation for a set of orbitals $\varphi_k$:}
\begin{equation}
  \I\hbar\pdiff{\varphi_k}{t}= \mat{H}\varphi_k + (\mat{S}_k - \I \mat{W}_k  \varphi_k)
  + \sum_l\lambda_{kl}\varphi_l.\label{eq:augmentedSchrodinger}
\end{equation}
Here, the optical potential $\I \mat{W}_k$ introduces dissipation, and a stochastic term $\mat{S}_k$ accounts for fluctuations.
These are linked through the fluctuation-dissipation theorem~\cite{Kubo1966}.
Lagrange multipliers $\lambda_{kl}$ are included to enforce the orthonormality $\braket{\varphi_k | \varphi_l} = \delta_{kl}$ at all times.

The dissipation term $\I \mat{W}_{k}$ poses a significant technical challenge because it breaks hermiticity, leading to non-unitary evolution in which the states $\varphi_{k}$ lose their orthonormality.
For small systems, this can be restored at each time step with e.g.\ Gram–Schmidt reorthogonalization, but this becomes prohibitively costly in large-scale three-dimensional fermionic simulations.
While specialized stochastic schemes such as quantum-jump or quantum-trajectory methods~\cite{Plenio} have been successfully applied in few-level or few-mode systems, they do not extend naturally to large Fermi systems, as they rely on non-Hermitian effective dynamics that require propagating full many-body wavefunctions: an approach that becomes computationally intractable in realistic 3D simulations.

To illustrate why modeling of \respch{damping} with a non-Hermitian operators quickly becomes computationally prohibitive, consider a single time step of the form
\begin{equation}
  \varphi_k(t+\Delta t) = e^{-\I \mat{H}_\tot\Delta t / \hbar}\varphi_k(t),
\end{equation}
applied to all $M$ evolved wave functions, where $\mat{H}_\tot$ includes both the Hermitian part of the Hamiltonian and a non-Hermitian term representing \respch{damping}.
If each $\varphi_k$ is stored as a vector of length $N$, the cost of applying the evolution operator is typically $O(MN\log N)$ when spatial derivatives are evaluated using spectral methods.
However, once the evolution is carried out with a non-Hermitian $\mat{H}_\tot$, the orthonormality of the wave functions is lost, requiring reorthogonalization at every step.
Using the Gram–Schmidt process, this operation scales as $O(M^2 N)$.
For large many-body systems, where $M \gg \log N$, this orthogonalization stage dominates the computational cost, often by orders of magnitude.

These issues can be avoided if \respch{the damping term is implemented efficiently while preserving orthogonality during evolution.}
This was the idea behind the \respch{so-called} \gls{LQF} algorithm~\cite{bulgac2013f, huangphd, Unitary_evolution_Bulgac}, which we extend here to the pairing channel.
The essence has two parts: 1) construct a time-dependent Hermitian potential that explicitly lowers the energy, and 2) restrict the potential to be diagonal (or local) in position (or momentum) space so that it can be efficiently implemented.
The simplest implementation~\cite{bulgac2013f} -- $U_{LQF}(\vect{r}, t) \propto \dot{n}(\vect{r}, t)$ -- is intuitive: the potential is repulsive (attractive) when the density is increasing (decreasing).
From the continuity equation, the potential thus slows the converging (diverging) flow, removing kinetic energy from the system.
Since the potential is diagonal (as opposed \respch{to} non-linear $U(\vect{r}, \vect{r}')$, which would require integration to apply), it is as easy to implement as an external or self-consistent potential, and its hermiticity means that evolution preserves the orthonormality of the $\varphi_k$.

\respch{Methods} that utilize hermitian evolution enable large-scale simulations of many-body systems, ranging from nuclear fission and heavy-ion collisions~\cite{Abdurrahman2024,Bulgac2024,Bulgac2025} to ultracold fermionic gases.
In the latter case, some of the most challenging simulations study quantum turbulence~\cite{PhysRevA.105.013304, Wlazlowski:10.1093, TowardsQT} which requires evolving $M = 10^5$ to $10^6$ 3D quasiparticle wave functions to fully capture the relevant physics.
Despite its efficient implementation, the \gls{LQF} algorithm can still stall for such large-scale problems, requiring prohibitive evolution times.
Intuitively, the \gls{LQF} algorithm can only remove energy from hydrodynamic flow [$n(\vect{r}, t)$ and $\vect{j}(\vect{r}, t)$], whereas large fermionic systems have many ``internal'' degrees of freedom associated with the quasiparticle wavefunctions $\varphi_k$.

In this work, we extend the \gls{LQF} algorithm to remove energy from collective excitations of the off-diagonal superfluid pairing fields.
We demonstrate that this significantly improves the \respch{cooling} efficiency in two demanding, but very different, superfluid systems: ultracold Fermi gases and dense nuclear matter in neutron stars.
The \respch{cooling} forces are implemented within the \gls{HFB} formalism, which is particularly well-suited for describing superfluid phenomena in fermionic systems.
Moreover, the proposed methodology is directly applicable in the context of \gls{TDDFT} studies for superfluids, which is a framework that is widely used in the nuclear physics and cold-atom communities, since the corresponding \gls{DFT} equations share the same formal structure as the \gls{HFB} equations~\cite{Bulgac2013,Bulgac2019,Schunck:2019,Magierski2019}.  

\section{Method}

We demonstrate our approach with the \gls{TDHFB} equations that are commonly used to model the dynamics of superfluid fermions.
These have the same formal structure as the \gls{TDDFT} equations used to study the \gls{UFG} (\gls{TDSLDA})~\cite{SLDA,Bulgac2013,Bulgac2019} and in nuclear physics~\cite{Schunck:2019,Magierski2019}.
The structure of these equations is (in units with $\hbar=1$)
\begin{gather}
  \label{eq:tdhfb}
  \I\pdiff{}{t} 
  \begin{pmatrix}
    u_{\mu}\\  
    v_{\mu}
  \end{pmatrix} =
\HHtd
  \begin{pmatrix}
    u_{\mu}\\
    v_{\mu}
  \end{pmatrix}
\end{gather}
where the quasiparticle states are given by the Bogoliubov amplitudes $\{u_{\mu}, v_{\mu}\}$ labeled by a set of quantum numbers $\mu$ (including e.g.\ energy and spin).
The total Hamiltonian effecting time-dependence $\HHtd$ consists of the intrinsic Hamiltonian $\HH$, and a dissipative term $\UU$
% and, if needed, a stochastic field  $\Stoch$ that accounts for fluctuations:
%
\respch{
\begin{equation}
  %\HHtd = \HH + \UU + \Stoch\,.
  \HHtd = \HH + \UU .
\end{equation}
}
\respch{
In this present paper, we focus on maximizing the effective damping due to $\UU$, which we refer to as the ``cooling potential''}. 
\respch{To ensure practical usability,}
we will construct $\UU$ to be diagonal in an appropriate representation, and Hermitian so that $\HHtd$ is Hermitian and the quasiparticle wavefunctions remain orthonormal throughout the evolution.

\subsection{General Form}
The general idea follows from expressing conserved quantities $Q = \Tr(\QQ \RR)$, like particle number $N$, in terms of the generalized density matrix $\RR$ \respch{[\cref{Hamiltonian_HFBtot}]}:
\begin{align}
  N &= \Tr( \NN \RR)\,,
  & \NN &= \begin{pmatrix}
    \mat{1} & \mat{0}\\
    \mat{0} & \mat{0}
  \end{pmatrix}\,.
\end{align}
These evolve under \cref{eq:tdhfb} as
\begin{subequations}
  \begin{align}
    \dot{\RR} &= \I[\RR,\HHtd],\\
    \dot{Q} &= \Tr\bigl(\QQ \dot{\RR}\bigr)= \I\Tr\bigl(\QQ[\RR,\HHtd]\bigr) = \I\Tr\bigl([\HHtd, \QQ]\RR]\bigr)\nonumber\\
            &= \I\Tr\bigl([\UU, \QQ]\RR]\bigr) = -\Tr\bigl(\I[\RR,\QQ]\UU]\bigr),
              \label{eq:Qdot}
  \end{align}
  where we assumed that $Q$ does not contain internal time dependence, and we have used the cyclic property of the trace combined with the fact that the conservation of the quantity $Q$ implied that $[\HH, \QQ] = 0$: hence, the dynamics under $\HHtd$ are generated entirely by the cooling potential $\UU$. 
In the density functional context, the dependence $E[ \Rdens]$ is generally non-linear and involves implicit time dependence via $\RR(t)$. However, the equations of motion still have the same form 
  \begin{align}
    \dot{E} &= \Tr\bigl(\tfrac{1}{2}\HH\dot{\RR}\bigr)
  \end{align}
where the Hamiltonian $\HH$ is \emph{defined} by varying $E[\mathscr{R}]$~\cite{RingSchuck}. The factor of $\tfrac{1}{2}$ appearing here accounts for the double-counting of particle-hole states in the Nambu-Gorkov formalism. In particular,
  \begin{align}
    \label{eq:ENdot}
    \dot{E} &= -\Tr\bigl(\I[\RR,\tfrac{1}{2}\HH]\UU\bigr)
    & \dot{N} &= -\Tr\bigl(\I[\RR,\NN]\UU]\bigr).
  \end{align}
\end{subequations}

The main point of this paper is that we can ensure cooling $\dot{E} \leq 0$ by evolving with a Hermitian cooling potential
\begin{equation}
  \label{eq:UU}
  \UU \propto \big(\I[\Rdens, \HH]\bigr)^\dagger
  = -\I[\HH,\Rdens]\,.
\end{equation}

\respch{
Note that $-\I[\HH,\Rdens] = \dot{\Rdens}$ is how the density matrix would evolve without the cooling potential $\UU$.
This leads to an intuitive picture similar to that in laser cooling of atoms or \textit{optical
molasses} \cite{CohenTannoudji1990}: the introduced cooling potential is large where the generalized density would change, thereby mitigating large changes and cooling the system.}

\glsreset{LQF}
The essence of the \respch{\gls{LQC}} algorithm is that local forms of this cooling potential can be efficiently implemented with state-of-the-art \gls{TDHFB}.
New in this paper is the use of cooling in the pairing channel, which can significantly improve cooling performance.
Additionally, we show how this can be used to adjust the particle number to simulate, e.g., particle loss, and to probe the properties of systems with different particle numbers / densities with a single time-dependent simulation -- a method we call a \respch{density scan}.

\subsection{TDHFB in Block Form}
It is convenient to explicitly rewrite the \gls{TDHFB} equations in block-diagonal form using the generalized density matrix $\Rdens$ and the \gls{TDHFB} Hamiltonian matrix $\mathscr{H}$~\cite{RingSchuck}, both of which are Hermitian:
\begin{subequations}
  \begin{align}\label{Hamiltonian_HFBtot}
    \RR
    &= 
      \begin{pmatrix}
        \mat{\rho} & \mat{\kappa} \\
        -\mat{\kappa}^* & \mat{1} - \mat{\rho}^*
      \end{pmatrix}\,,
    & \HH
    &=
      \begin{pmatrix}
        \h & \D \\
        -\D^* & -\h^*
      \end{pmatrix}\,.
  \end{align}
  The \gls{TDHFB} Hamiltonian matrix generates time-evolution in $\Rdens$ through the anti-Hermitian commutator:
  \begin{equation}
    \label{eq:iRdot}
    \I \dot{\RR} = 
    \big[\HH, \Rdens\big] =
    \begin{pmatrix}
      \A & \B \\
      \B^* & \A^* 
    \end{pmatrix}\,.
  \end{equation}
  The blocks of $\Rdens$ contain the normal quasiparticle density matrix $\mat{\rho}$, and the anomalous quasiparticle density matrix $\mat{\kappa}$, which can be expressed as expectation values of the fermionic annihilation and creation operators $\op{a}_\mu$ and $\op{a}^\dagger_\mu$:
  \begin{align}
    \label{eq:rho}
    \rho_{\mu \nu} &= \braket{\op{a}^\dagger_\nu \op{a}_\mu}
    & \kappa_{\mu \nu} &= \braket{\op{a}_\nu \op{a}_\mu}
  \end{align}
  and the blocks of $\HH$ contain the single-particle Hamiltonian $\h$ and pairing potential $\D$.
  We have also expressed the blocks structure of $\I\dot{\Rdens}$ in terms of the matrices $\A$ and $\B$.
  Hermiticity of $\Rdens$ and $\HH$ and the anti-commutation of the fermionic operators require the following symmetries
  \begin{align}
    \mat{\rho} &= \mat{\rho}^\dagger, & \mat{\kappa} &= -\mat{\kappa}^{T}\,,\\
    \h &= \h^\dagger, & \D &= -\D^T\,,\label{eq:Dskew}\\
    \A &= -\A^\dagger, & \B &= -\B^T\,.\label{eq:Askew}
  \end{align}
\end{subequations}
In block form, the evolution of the density matrix becomes
\begin{subequations}
  \label{eq:ABtot}
  \begin{align}
    \I\dot{\mat{\rho}} &= \A = \big[\h, \mat{\rho}\big]
                         - \D \mat{\kappa}^* + \mat{\kappa} \D^*\,, \\
    \I\dot{\mat{\kappa}} &= \B = \h \mat{\kappa} + \mat{\kappa} \h^*
                           + \D (\mat{1} - \mat{\rho}^*) - \mat{\rho}\D\,,
  \end{align}
\end{subequations}
and the optimal Hermitian cooling potential \cref{eq:UU} is:
\begin{equation}
  \UU \propto
  -\I[\HH, \RR]
  = 
  -\I
  \begin{pmatrix}
    \A & \B \\
    \B^* & \A^* 
  \end{pmatrix}\,.
\end{equation}
While guaranteed to lower the energy, this may not provide enough flexibility. \respch{In order to simplify calculations and improve the speed of cooling, we evolve with:}
\begin{gather}
  \label{eq:Ufricf}
  \UU =
  -\I
  \begin{pmatrix}
    \Atd & \Btd \\
    \Btd^* & \Atd^* 
  \end{pmatrix}\,.
\end{gather}
With this evolution, from \cref{eq:ENdot,eq:Askew,eq:iRdot} we have
\begin{subequations}
  \label{eq:Cooling}
  \begin{align}
    \dot{E} &= \tfrac{1}{2}\Tr\Bigg(
    \overbrace{
      \begin{pmatrix}
        \A & \B \\
        \B^* & \A^* 
      \end{pmatrix}}^{-[\Rdens, \HH]}
    \overbrace{
      \begin{pmatrix}
        \Atd & \Btd \\
        \Btd^* & \Atd^* 
      \end{pmatrix}}^{\I\UU}
    \Bigg)\\
    %= \Tr(\A\Atd + \A^*\Atd^* + \B\Btd^* + \B^*\Btd)
    &= -\Re\Tr(\A^\dagger\Atd + \B^\dagger\Btd).
  \end{align}
\end{subequations}
New to this paper, we note that from \cref{eq:ENdot},
\begin{gather}
  \dot{N} = -\Tr\Bigl(\I[\Rdens, \NN]\UU\Bigr)
  = -2\Re\Tr(\mat{\kappa}^\dagger\Btd)\,,
\end{gather}
so we can also use $\Btd$ to adjust the particle number.

Summarizing, we have
\begin{subequations}
  \label{eq:dotEN}
  \begin{align}
    \dot{E} = - &\Re\Tr(\B^\dagger\Btd) - \Re\Tr(\A^\dagger\Atd)\,,\\
    \dot{N} = - 2&\Re\Tr(\mat{\kappa}^\dagger\Btd)\,.
  \end{align}
\end{subequations}
Thus, we can use $\Atd$ and $\Btd$ to adjust the energy of the system, and, by appropriately choosing $\Btd$, we can also adjust the particle number.
(See \cref{appendix-A} for details.)

\subsection{Local Quantum \respch{Cooling}}
For practical purposes, computing with the full cooling matrix $\UU$ can be computationally expensive.
The second aspect of the \respch{\gls{LQC}} algorithm is to note that we still have guaranteed cooling if we restrict the form of $\Atd$ and $\Btd$.
For example, we can choose
\begin{subequations}
  \label{eq:alphabeta}
  \begin{align}
    [\Atd]_{mn} &= \tilde{\alpha}_{mn}A_{mn},
    & [\Btd]_{mn} &= \tilde{\beta}_{mn}B_{mn},
  \end{align}
  with real non-negative coefficients $\tilde{\alpha}_{mn} \geq 0$ and $\tilde{\beta}_{mn} \geq 0$ so that the cooling property is preserved:
  \begin{align}
    \dot{E} &= -\sum_{mn} \left(\tilde{\beta}_{mn}\abs{B_{mn}}^2 + \tilde{\alpha}_{mn}\abs{A_{mn}}^2\right)
              \leq 0\,.
  \end{align}
\end{subequations}
Before we show how \respch{\gls{LQC}} in the pairing channel with non-zero $\tilde{\beta}_{mn}$ speeds cooling, we first review the coordinate-space expressions for densities and the single-particle Hamiltonian.

\subsection{HFB Equations in Coordinate Space}
The spatial dependence of the density matrix $\mat{\rho}$ for a single species can be expressed in terms of a $2\times 2$ block structure from the spin degrees of freedom $\sigma \in \{\ua, \da\}$:
\begin{equation}
  \braket{\vect{r}|\mat{\rho}|\vect{r}'} = 
  \begin{pmatrix}
    \rho_{\ua\ua}(\vect{r}, \vect{r}^\prime) & \rho_{\ua\da}(\vect{r}, \vect{r}^\prime)\\
    \rho_{\da\ua}(\vect{r}, \vect{r}^\prime) & \rho_{\da\da}(\vect{r}, \vect{r}^\prime)
  \end{pmatrix},
\end{equation}
where each submatrix is expressed in terms of the quasiparticle Bogoliubov amplitudes appearing in \cref{eq:tdhfb}:
\begin{align}
  \label{eq:rhoss}
  \rho_{\sigma\sigma'}(\vect{r}, \vect{r}') = \sum_n v_{n\sigma}^*(\vect{r})v_{n\sigma'}(\vect{r}').
\end{align}
An analogous structure appears for the anomalous density
\begin{align}
  \braket{\vect{r}|\mat{\kappa}_{\sigma\sigma'}|\vect{r}'} = \sum_n v_{n\sigma}^*(\vect{r})u_{n\sigma'}(\vect{r}').
\end{align}

The single-particle Hamiltonian has the general form
\begin{equation}
  \mat{h} = 
  \begin{pmatrix}
    \mat{h}_{\ua\ua} & \mat{h}_{\ua\da}\\
    \mat{h}_{\da\ua} & \mat{h}_{\da\da}
  \end{pmatrix},
\end{equation}
where the off-diagonal terms $\mat{h}_{\ua\da} = \smash{\mat{h}_{\da\ua}^\dagger}$ arise from spin-coupling interactions, such as the spin–orbit term in nuclear systems.
For clarity, we will neglect these ($\mat{h}_{\ua\da}=0$), but our results can be easily extended to include these terms if needed.
Note that in both of our applications, neglecting these is a good physical approximation:
In ultracold atoms, the coupling is often absent (though it can be engineered~\cite{dalibard2011colloquium});
In neutron stars, the neutron density gradients are sufficiently small that the spin–orbit interaction remains negligible.

For non-relativistic systems, the remaining blocks are
\begin{equation}\label{eq:hss}
  \braket{\vect{r}|\mat{h}_{\sigma\sigma}|\vect{r}'} = \delta(\vect{r}-\vect{r}')\left(\frac{-\nabla^2}{2m} + U_{\sigma}(\vect{r})\right),
\end{equation}
where the first term is the kinetic energy of a particle of mass $m$, and $U_\sigma(\vect{r})$ is the mean-field potential, which may also include the chemical potential.
Hereafter, we suppress the $\delta(\vect{r} - \vect{r}')$ factor, using the notation $U(\vect{r})$ with a single coordinate $\vect{r}$ in parentheses to denote an operator for which the matrix representation in coordinate space is diagonal.

The same structure appears in the pairing field.
Assuming pairing only between opposite spins (i.e., neglecting $P$-wave pairing, etc.), the pairing matrix is
\begin{subequations}
  \label{eq:Deltalocal}
  \begin{equation}
    \label{eq:Deltamat}
    \D = 
    \begin{pmatrix}
      0 & \Delta_{\ua\da}(\vect{r})\\
      \Delta_{\da\ua}(\vect{r}) & 0
    \end{pmatrix}
    =
    \begin{pmatrix}
      0 & \Delta(\vect{r})\\
      -\Delta(\vect{r}) & 0
    \end{pmatrix},
  \end{equation}
  and we can use the properties \cref{eq:Dskew} of $\D$ to express $\Delta_{\ua\da}(\vect{r}) = -\Delta_{\da\ua}(\vect{r}) = \Delta(\vect{r})$ in terms of a single function, for which we drop the spin indices.
  This is expressed in terms of the anomalous density
  \begin{equation}
    \label{eq:geff}
    \Delta(\vect{r}) = -g_{\textrm{eff}}(\vect{r})\kappa_{c}(\vect{r})\,,
  \end{equation}
  where $g_{\textrm{eff}}$ is an effective coupling constant.
  For the attractive interactions we consider here (needed to sustain superfluid pairing), $g_{\text{eff}} < 0$, hence $\Delta(\vect{r})$ and $\kappa(\vect{r})$ have the same phase and sign:
  \begin{gather}
    \Delta(\vect{r}) = \abs{\Delta(\vect{r})}e^{\I\theta_{\kappa}}\,, \qquad
    \kappa(\vect{r}) = \abs{\kappa(\vect{r})}e^{\I\theta_{\kappa}}\,.
  \end{gather}
  To render this term local in position was a major technical advancement~\cite{Bulgac:2002, Bulgac:2007, Bulgac:2012, Boulet:2022}, which is non-trivial since the anomalous density $\kappa(\vect{r}) = \lim_{\vect{r}'\rightarrow\vect{r}}\kappa(\vect{r}, \vect{r}')$ formally diverges.
  The quantities in \cref{eq:Deltalocal} reflect this regularization.
\end{subequations}

Modern time-dependent \gls{HF} and \gls{HFB} solvers~\cite{maruhn2014sky3d,Jin2021,pecak2024WBSkMeff,WSLDAToolkit,Abhishek2024,Marevi2024,Yoshimura2024} exploit this ability to confine all non-locality in $\HH$ to derivatives ($\vect{\nabla}$, $\nabla^2$, etc.) by using efficient pseudo-spectral methods based on the \gls{FFT}, which efficiently transforms the wavefunctions from position to momentum space and back.
Derivatives are thus computed by point-wise (vector-vector) multiplication in momentum space, while all other operations are point-wise multiplications in position space.

To maintain this efficiency, one should use a cooling potential $\UU$ that is local in position (or momentum), corresponding to diagonal coefficients $\alpha_{ab}, \beta_{ab} \propto \delta_{ab}$ in the appropriate basis in \cref{eq:alphabeta}.
Here we consider only cooling potentials that are diagonal in position space~\footnote{Cooling in momentum space has been considered by P.~Danielewicz (private communication) and in \cite{huangphd}, but has not been widely used yet in the literature.}.
With these approximations, we implement the \respch{\gls{LQC}} algorithm by augmenting the time-evolution with three functions of position, $U_{\ua}\diss$, $U_{\da}\diss$, and $\Ddiss$:
\begin{subequations}
  \begin{align}
    \left(\htd\right)_{\sigma\sigma}
    &= \mat{h}_{\sigma\sigma} + \Udiss_{\sigma}(\vect{r})\,,\label{eq:hss1}\\
    \left(\Delta_\td\right)_{\ua\da} \equiv \Delta_\td(\vect{r})
    &= \Delta(\vect{r})+\Ddiss(\vect{r})\,.\label{eq:Deltalocal1}
  \end{align}
\end{subequations}
For maximal cooling, we will generally take all coefficients $\alpha_{aa} = \alpha$ to be the same: i.e.
\begin{gather}
  \Udiss_{\sigma}(\vect{r}) = -\I\alpha \diag(\mat{A}_{\sigma\sigma}),
\end{gather}
but note that one can further exploit this locality to implement localized cooling with a function $\alpha(\vect{r})$ that is selectively non-zero -- e.g., near the boundaries of a simulation.

\subsection{\respch{Cooling} mean-field \texorpdfstring{$\Udiss$}{U}}
We first consider the case where the proportionality coefficients in \cref{eq:Ufricf} are chosen to retain only the diagonal terms:
\begin{equation}
  \UU_U = -\I \tilde{\alpha}
  \begin{pmatrix}
    \diag(\A) & \mat{0} \\ 
    \mat{0}         & \diag(\A)^*
  \end{pmatrix}, 
\end{equation}
with $\tilde{\alpha}>0$ controlling the strength of the \respch{cooling} potential.  
The diagonal part can be written using \cref{eq:ABtot}~\footnote{The term $\diag\bigl(\D\mat{\kappa}^* - \mat{\kappa}\D^*\bigr)
  = \diag\bigl(\D\mat{\kappa}^* - (\D\mat{\kappa}^*\bigr)^\dagger\bigl) = 0$ vanishes because $\Delta(\vect{r})$ and $\kappa(\vect{r})$ have the same phase~\cref{eq:geff}.}
\begin{align}
  \diag(\A) & = \diag\bigl([\h, \mat{\rho}] - \D\mat{\kappa}^* + \mat{\kappa}\D^*\bigr)\nonumber\\
            & = \diag\bigl([\h, \mat{\rho}]\bigr),
\end{align}
Expanding explicitly in spin gives
\begin{align}
    \diag(\A) = 
    \begin{pmatrix}
     \diag\Bigl(\bigl[\h_{\ua\ua}, \mat{\rho}_{\ua\ua}\big]\Bigr) & \mat{0}\\
      \mat{0} & \diag\Bigl(\big[ \h_{\da\da}, \mat{\rho}_{\da\da} \big]\Bigr)
    \end{pmatrix},
\end{align}
where we have used our approximation $\h_{\ua\da}=\mat{0}$.
Spin-off-diagonal components can be included as well if used.
The resulting \respch{cooling} mean-field potential takes the form
\begin{equation}
  \Udiss_{\sigma}(\vect{r}) \equiv -\I \tilde{\alpha} \diag\Bigl(\big[\h_{\sigma\sigma}, \mat{\rho}_{\sigma\sigma}\big]\Bigr),
\end{equation}
which can be evaluated explicitly by inserting $\h_{\sigma\sigma}$ from \cref{eq:hss} and $\rho$ from \cref{eq:rhoss}.
Omitting the spatial coordinate for brevity, we obtain
\begin{align}
  \label{eq:comm_hrho}
  \Udiss_{\sigma} &= -\I \tilde{\alpha}\,\diag\Biggl(\biggl[\frac{-\nabla^2}{2m}, \rho_{\sigma\sigma}\biggr]\Biggr)\nonumber\\
                  &= \frac{\I \tilde{\alpha}}{2m}\sum_n \biggl(\Bigl(\nabla^2 v_{n,\sigma}^*\Bigr)v_{n,\sigma} - v_{n,\sigma}^*\Bigl(\nabla^2 v_{n,\sigma}\Bigr)\biggr)\,\nonumber\\
                  &= \frac{\tilde{\alpha}}{m}\sum_n \Im\biggl(v_{n,\sigma}^*\Bigl(\nabla^2 v_{n,\sigma}\Bigr)\biggr),
\end{align}
where all $v_{n,\sigma}$ are evaluated at the same spatial point $\vect{r}$.
In deriving this result, we used that the mean-field potential $U_\sigma$ is diagonal, and that the commutator of a diagonal matrix with any matrix has vanishing diagonal elements.
Introducing the current density
\begin{equation}
  \vect{j}_\sigma(\vect{r}) = -\sum_n \Im\bigl(v_{n,\sigma}^*(\vect{r})\vect{\nabla} v_{n,\sigma}(\vect{r})\bigr),
\end{equation}
the mean-field \respch{cooling} potential thus takes the compact form
\begin{equation}
  \Udiss_\sigma(\vect{r}) = -\frac{\tilde{\alpha}}{m}\vect{\nabla}\cdot \vect{j}_\sigma(\vect{r}).
\end{equation}
It is convenient to introduce a dimensionless \respch{cooling} constant $\alpha = \rho_0\tilde{\alpha}$, where $\rho_0$ is a characteristic density of the system (e.g., the nuclear saturation density for nuclear systems).  
The single-particle Hamiltonian for the time evolution then reads
\begin{equation}
  \label{eq:cooling_alpha}
  \left(\htd\right)_{\sigma\sigma} = \h_{\sigma\sigma} - \frac{\alpha}{m\rho_0}\underbrace{\vect{\nabla}\cdot \vect{j}_\sigma(\vect{r})}_{-\dot{\rho}_{\sigma}(\vect{r})}
\end{equation}
and generates \respch{cooling} dynamics whenever nonzero currents $\vect{j}_\sigma$ are present (i.e., where the density would change $\dot{\rho}$ under evolution without cooling).
This is the original form of \respch{\gls{LQC}} introduced in~\cite{bulgac2013f} as a force opposing irrotational currents in the system, thereby damping them.

\subsection{\respch{Cooling} Pairing Field \texorpdfstring{$\Ddiss$}{Delta}}
One can also construct a \respch{cooling} pairing field.
To retain the advantages of a local formulation, we assume it has the form
\begin{equation}
  \label{u_delta}
  \UU_\Delta = -\I
  \begin{pmatrix}
    \mat{0}          & \Btd \\ 
    \Btd^*  & \mat{0}
  \end{pmatrix}, 
\end{equation}
where the matrix $\Btd$ has the same structure as the pairing field matrix \cref{eq:Deltamat}, namely
\begin{equation}
  \Btd =
  \tilde{\beta}(\vect{r})
  \begin{pmatrix}
    \mat{0} & B(\vect{r})\\
    -B(\vect{r}) & \mat{0}
  \end{pmatrix}.
\end{equation}
Here, $\tilde{\beta}(\vect{r})$ is a position-dependent proportionality coefficient.
The corresponding pairing cooling potential is then
\begin{equation}
  \label{eq:Ddiss}
  \Ddiss(\vect{r})\equiv -\I\tilde{\beta}(\vect{r})B(\vect{r}).
\end{equation}
where $B(\vect{r})$ denotes the upper right component of the diagonal part of \cref{eq:ABtot}.
In contrast with $\tilde{\alpha}$, the coefficient $\tilde{\beta}(\vect{r})$ needs some spatial dependence because $B(\vect{r})$ is formally divergent.
Thus, we expect at least $\tilde{\beta}(\vect{r}) \sim \abs{g_{\text{eff}}(\vect{r})}$ to render finite the anomalous cooling potential as in \cref{eq:geff}.
With this in mind, the functional form of $B(\vect{r})$ can be expressed as
\begin{subequations}
  \begin{align}
    \label{eq:Bdiss0}
    B(\vect{r}) &=
    \diag\bigl(\h_{\ua\ua}\mat{\kappa} + \mat{\kappa}\h^*_{\da\da} + \D(\mat{1} - \mat{\rho}^*) -\rho \D\bigr),\\
                &= \Delta(\vect{r})\Biggl[1-\rho_{\ua\ua}(\vect{r})-\rho_{\da\da}(\vect{r})-\frac{U_\ua(\vect{r})}{g_{\textrm{eff}}(\vect{r})} - \frac{U_\da(\vect{r})}{g_{\textrm{eff}}(\vect{r})}\Biggr]\nonumber\\
                &   -\frac{1}{2m}\sum_n\left[u_{n\da}(\vect{r})\nabla^2v_{n\ua}^*(\vect{r}) + v_{n\ua}^*(\vect{r})\nabla^2u_{n\da}(\vect{r}) \right],
                  \label{eq:Bdiss}
  \end{align}
\end{subequations}
where we have used the locality of the pairing field, \cref{eq:Deltalocal}.

\subsection{Particle Number Conservation}
From \cref{eq:dotEN}, the energy and particle number evolve as
\begin{subequations}
  \begin{align}
    \dot{E} = - &\Im\Tr\bigl(B^*(\vect{r})\Delta\diss(\vect{r})\bigr) - \Re\Tr(\A^\dagger\Atd)\,,\\
    \dot{N} = - 2&\Im\Tr\bigl(\kappa^*(\vect{r})\Delta\diss(\vect{r})\bigr)\,.
  \end{align}
\end{subequations}
Thus, to if $\kappa^*(\vect{r}) \Delta\diss(\vect{r})$ is real, particle number will be conserved.
For the intrinsic pairing field this holds automatically since $\D(\vect{r})$ and $\kappa(\vect{r})$ share the same phase [\cref{eq:geff}]
\begin{subequations}
  \begin{gather}
    \theta_{\kappa}(\vect{r}) = \arg\bigl(\kappa(\vect{r})\bigr)\,,\\
    \begin{aligned}
      \kappa(\vect{r})
      &= \abs{\kappa(\vect{r})}e^{\I\theta_\kappa(\vect{r})}\,,
      & \Delta(\vect{r})
      &= \abs{\Delta(\vect{r})}e^{\I\theta_\kappa(\vect{r})}\,.
    \end{aligned}
  \end{gather}
\end{subequations}
To see how we can do this while maintaining cooling, let (suppressing the dependence on $\vect{r}$ to simplify the notation)
\begin{align}
  \label{eq:abcd}
  B &= (a - \I b)e^{\I\theta_{\kappa}}\,,
  & \Delta\diss &= (c + \I d)e^{\I\theta_{\kappa}}\,.
\end{align}
Then, ignoring the cooling from $-\Re\Tr(\A^\dagger\Atd)$, we have
\begin{subequations}
  \begin{align}
    \dot{E} = - &\Im\Tr\bigl((a+\I b)(c+\I d)\bigr) = - \Tr(ad + bc)\,,\\
    \dot{N} = - 2&\Im\Tr\bigl(\abs{\kappa}(c+\I d)\bigr) = - 2\Tr(\abs{\kappa}d)\,.
  \end{align}
\end{subequations}
Thus, if we use the cooling potential \cref{eq:Ddiss} discussed above: $\Delta\diss = -\I\tilde{\beta}B$, then $c = \tilde{\beta}b$ and $d = \tilde{\beta} a$, so $\dot{E} = - \Tr\bigl(\tilde{\beta}(a^2 + b^2)\bigr)$, cooling as expected, but $\dot{N} = - 2\Tr(\tilde{\beta}\abs{\kappa}a)$.
Thus, the particle number will also change if $a$ is non-zero.
This is a peculiar feature of superfluidity and the mean-field description of the spontaneously broken phase symmetry.

To ensure cooling while preserving particle number, we must set $d = 0$ and choose $c$ to have the same sign as $b$ so that $bc \geq 0$.
To do this, we note that the term from the first line of \cref{eq:Bdiss} contributes only to $a$:
\begin{align}\label{eq:Dnonconserv}
  a = \abs{\Delta}\Biggl(1 - \rho_{\ua\ua} - \rho_{\da\da}
  - \frac{U_\ua}{g_{\textrm{eff}}}
  - \frac{U_\da}{g_{\textrm{eff}}}\Biggr)
  + \cdots,\,
\end{align}
while the term in the second line of \cref{eq:Bdiss} contributes to both
\begin{subequations}
  \label{eq:diagB}
  \begin{align}
    B' &= -\frac{1}{2m}\sum_n\left[
         u_{n\da}\nabla^2v_{n\ua}^* + v_{n\ua}^*\nabla^2u_{n\da}
         \right]\,,\\
       &= \abs{B'}e^{\I\theta_{B'}}
         = \abs{B'}\underbrace{(\cos \phi + \I \sin\phi)}_{e^{\I\phi}}e^{\I\theta_{\kappa}}\,,
  \end{align}
  where $\phi = \theta_{B'} - \theta_{\kappa}$ is the difference in phases between $B'$ and $\kappa$. 
\end{subequations}
Thus:
\begin{align}
  b &= -\abs{B'}\sin\phi\,,
\end{align}
and we can ensure cooling without changing the particle number by setting $d=0$ and $c \propto b$:
\begin{subequations}
  \begin{align}
    \Delta\diss(\vect{r})
    &= -\tilde{\beta}(\vect{r})\abs{B'(\vect{r})}\sin\phi(\vect{r}) e^{\I\theta_\kappa(\vect{r})}\,,\\
    &= -\beta\sin\bigl(\theta_{B'}(\vect{r}) - \theta_{\kappa}(\vect{r})\bigr)\Delta(\vect{r}).
  \end{align}
\end{subequations}
In the last line, we have introduced a dimensionless constant $\beta > 0$ to keep $\Delta\diss$ the same magnitude as the pairing field:
\begin{gather}
  \label{eq:tilde-beta}
  \tilde{\beta}(\vect{r}) = \beta\frac{\abs{\Delta(\vect{r})}}{\abs{B'(\vect{r})}}.
\end{gather}

In summary, the pairing potential \cref{eq:Deltalocal1} to be used in the evolution reads
\begin{align}
  \label{eq:Deltadiss1}
  \Delta_\td(\vect{r}) = \Delta(\vect{r})\Bigl(
    1 - \beta \sin\bigl(\theta_{B'}(\vect{r}) - \theta_\kappa(\vect{r})\bigr)
  \Bigr), 
\end{align}
where we used that $\Delta=\abs{\Delta}e^{\I\theta_\kappa}$.
This construction generates \respch{cooling} dynamics while preserving locality and conserving particle number.
A \respch{cooling} potential of this form minimizes excitations in the pairing field.
Moreover, its simple structure makes it straightforward to implement in practice.

\subsection{Adjusting the Particle Number}
\label{ssec:Controlling-particle-number}
To maintain particle-number conservation, we set $d = 0$ in \cref{eq:abcd} where $\Delta\diss = (c+\I d)e^{\I\theta_\kappa}$.
We now consider non-zero $d$ so that we can adjust the particle number as we evolve.
This has several applications: for example, simulating particle loss in ultracold atomic gases, such as those induced by three-body recombination or thermal effects~\cite{3BL_BEC1,3BL_BEC2}.
Another application is to guide the particle number toward a desired value while preparing initial states.
A third application -- which we call a density scan -- is to slowly vary the particle number in a system so that a single time-dependent simulation can access properties of many states with different particle numbers, achieving what would take conventional approaches many expensive simulation runs.

The approach is simply to add an imaginary piece to \cref{eq:Deltadiss1}:
\begin{align}
  \label{eq:Deltadiss2}
  \Delta_\td(\vect{r}) =  \Delta(\vect{r})\Bigl(
    1 - \beta\sin\bigl(\theta_{B'}(\vect{r}) - \theta_\kappa(\vect{r})\bigr)
    + \I\tilde{\gamma}
  \Bigr),
\end{align}
where the parameter $\beta$ controls the strength of the particle-conserving term, while $\tilde{\gamma}$ regulates the change in $N$:
\begin{equation}
  \dot{N} = -2\tilde{\gamma}\,\Tr\bigl(\abs{\Delta(\vect{r})}\,\abs{\kappa(\vect{r})}\bigr). 
\end{equation}
For example, to guide the particle number to a desired value $\Nreq$, one can use
\begin{equation}
  \label{eq:tilde-gamma}
  \tilde{\gamma} = \gamma\frac{N(t)-\Nreq(t)}{\Nreq(t)},
\end{equation}
so that the external pairing potential drives the system toward the requested particle number $\Nreq$, which may itself be time dependent.

Note: arbitrarily choosing $\tilde{\gamma}$ to adjust the particle number might spoil the cooling property $\dot{E} = -\Tr(ad + bc) - \Re\Tr(\A^\dagger\Atd)$.
If simultaneous cooling is required, then one should ensure that cooling from the other terms is sufficient, or choose $\tilde{\gamma}(\vect{r})$ so that $ad$ remains positive where
\begin{subequations}
  \begin{align}
    a &= \abs{\Delta}\Biggl(1 - \rho_{\ua\ua} - \rho_{\da\da}
        - \frac{U_\ua}{g_{\textrm{eff}}}
        - \frac{U_\da}{g_{\textrm{eff}}}\Biggr)
        + \abs{B'}\cos\phi,\,\\
    d &= \frac{\tilde{\gamma}(\vect{r})}{\abs{\Delta(\vect{r})}}\,,
  \end{align}    
\end{subequations}
which might require adjusting $\tilde{g}(\vect{r})$ spatially to capture the appropriate signs.
If this is insufficient, the more complicated formalism discussed in \cref{appendix-A} can be used.

\section{Tests of \respch{cooling} dynamics}

To validate our methodology, we perform a series of numerical simulations designed to test the robustness and accuracy of the proposed \respch{cooling} terms.
These simulations are performed on systems with either known outcomes or well-understood physical behavior.
The first set of tests examines whether \respch{cooling} dynamics can serve as an efficient tool for ground-state preparation.
For this purpose, we initialize time-dependent runs from states close to the ground state and observe how the system relaxes toward it.  
In the second set of tests, we apply an external perturbation to a uniform system and then study the subsequent evolution both with and without cooling.
Together, these controlled experiments provide a systematic assessment of the performance and convergence properties of the proposed methods.

For both test cases, and for later applications with spin-imbalanced ultracold Fermi gases, we use the \gls{ASLDA}~\cite{PhysRevLett.101.215301,Bulgac:2012}, which defines the intrinsic energy density functional as
\begin{gather}
  \label{eq:sldae-functional-CM}
  \mathscr{E}(\vect{r})
  =
  \frac{\tau(\vect{r})}{2m} + \frac{3}{5}B(p) \rho(\vect{r}) \eF(\vect{r}) + \frac{C(p)}{\rho(\vect{r})^{1/3}} |\kappa_{\ua\da}(\vect{r})|^2\,,
\end{gather}
where $\rho(\vect{r})=\rho_{\ua\ua}(\vect{r})+\rho_{\da\da}(\vect{r})$ is the total density, $\eF(\vect{r})=\kF(\vect{r})^2/2m$ is the local Fermi energy, and $\kF(\vect{r})=\bigl[3\pi^2\rho(\vect{r})\bigr]^{1/3}$ is the local Fermi momentum.
The kinetic energy density is defined as
 \begin{equation}\label{eq:kinetic-density}
   \tau(\vect{r}) = \sum_{n,\sigma}\abs{\vect{\nabla} v_{n,\sigma}(\vect{r})}^2\,,
 \end{equation}
and the coefficients $B(p)$ and $C(p)$ depend implicitly on the local spin polarization 
\begin{equation}
  p \respch{(\vect{r})}= \frac{\rho_{\ua\ua}(\vect{r})-\rho_{\da\da}(\vect{r})}{\rho_{\ua\ua}(\vect{r})+\rho_{\da\da}(\vect{r})}\,,
\end{equation}
with explicit forms given in~\cite{Bulgac:2012}.
These coefficients are tuned to reproduce quantum Monte Carlo results for the unitary Fermi gas at varying spin polarizations.

\begin{figure*}[t]
  \centering
  \includegraphics[width=\textwidth]{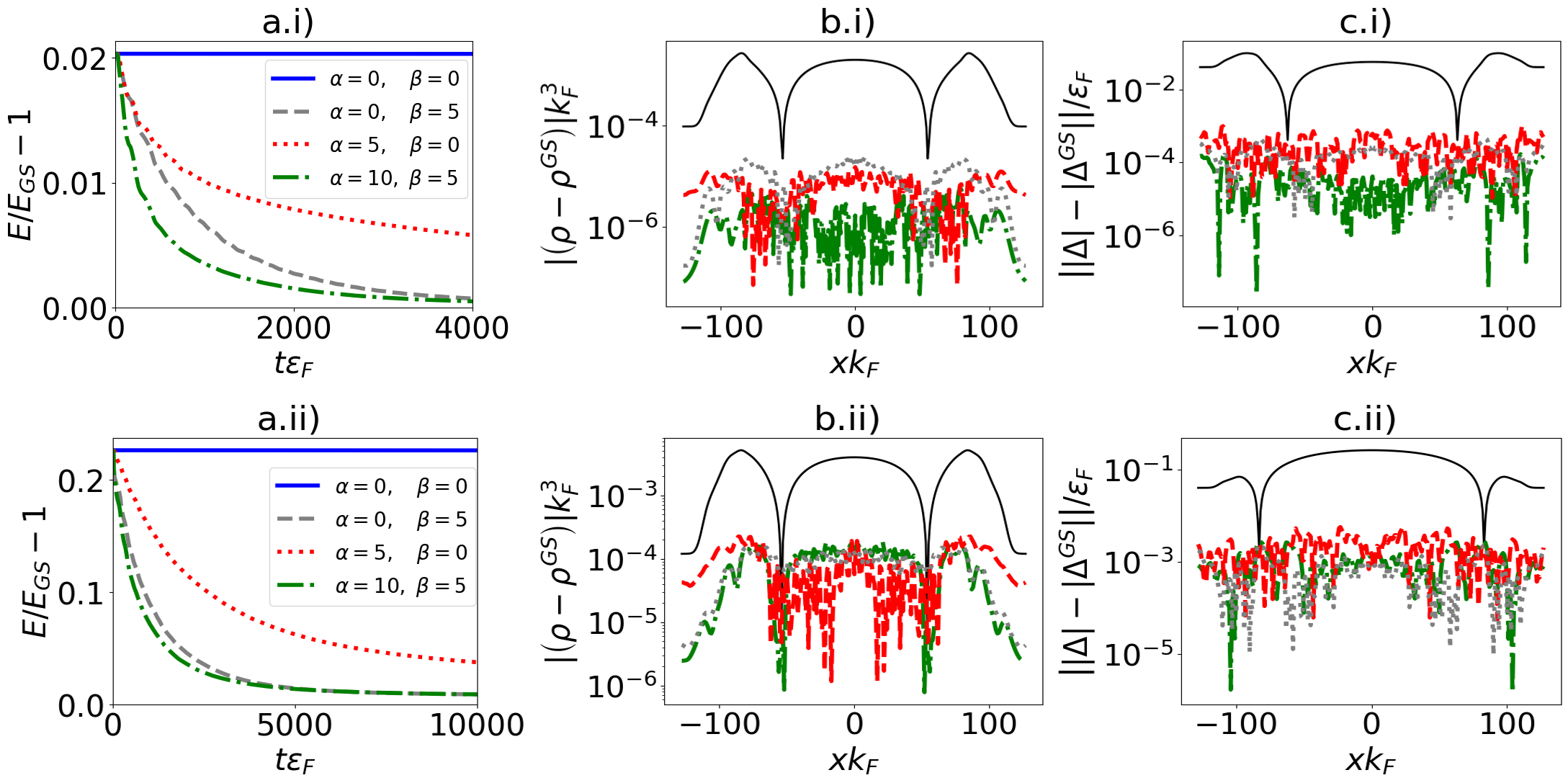}
  \caption{Particle-conserving time evolution ($\gamma = 0$) of a state under different \respch{cooling} parameter values.
    Rows (i) and (ii) correspond to initial states with excitation energies of $2\%$ and $25\%$, respectively.
    Column (a) shows the excitation energy as a function of time.
    Column (b) displays the difference between the ground-state density and the density of the evolved state at the end of the simulation; colors match the legend in column (a), and the solid thin black line indicates the difference between the ground-state density and the initial density at $t = 0$.
    Column (c) is analogous to column (b), but shows differences in the absolute value of the pairing field instead of density.
    \label{fig:energy_PN_1d}}
\end{figure*}

The total intrinsic energy is obtained by integrating the energy density functional over space,
\begin{equation}
  E = \int \mathscr{E}(\vect{r})\, d\vect{r}\,,
\end{equation}
and the corresponding mean-field and pairing potentials are defined through functional derivatives,
\begin{align}
  U_{\sigma}(\vect{r}) &= \frac{\delta E}{\delta \rho_{\sigma\sigma}}\,,
  &  \Delta(\vect{r}) &= - \frac{\delta E}{\delta \kappa^*}\,.
\end{align}
These enter the evolution equations through the single-particle Hamiltonian \cref{eq:hss1} and the pairing potential $\Delta(\vect{r})$ \cref{eq:Deltalocal1}.
In the spirit of \gls{DFT}, the single-particle Hamiltonian can be augmented with an external potential, 
\begin{equation}
  \h_{\sigma\sigma} \rightarrow \h_{\sigma\sigma} + V_{\textrm{ext}}(\vect{r})\,.
\end{equation}

We solve these equations with the \gls{WSLDA} Toolkit~\cite{WSLDAToolkit}, which provides solvers for both static and time-dependent equations with the same structure as the \gls{HFB} framework.
The simulations are performed on a three-dimensional lattice of size $N_x\times N_y\times N_z$ with uniform spacing $\d{x}=\d{y}=\d{z}=1$, which sets our length scale.

In the following tests, 1D or 2D refers to systems that are spatially uniform in one or two directions, respectively; i.e., all simulations are performed with a 3D equation of state.
Concretely, in the 1D case, the densities depend only on a single coordinate, e.g.,
\begin{equation}
  \rho_{\sigma\sigma}(\vect{r}) \rightarrow \rho_{\sigma\sigma}(x)\,,
\end{equation}
while in the 2D case, they depend on two coordinates,
\begin{equation}
  \rho_{\sigma\sigma}(\vect{r}) \rightarrow \rho_{\sigma\sigma}(x,y)\,.
\end{equation}
As a result, the Bogoliubov amplitudes $\{u_{n,\sigma}, v_{n,\sigma}\}$ reduce to plane waves along the remaining directions.
For instance, in the 2D case, they take the structure
\begin{equation}
  \begin{pmatrix}
    u_{n,\sigma}(\vect{r})\\  
    v_{n,\sigma}(\vect{r})
  \end{pmatrix}
  =
  \begin{pmatrix}
    u_{n,\sigma}(x,y)\\  
    v_{n,\sigma}(x,y)
  \end{pmatrix}
  e^{\I k_z z},  
\end{equation}
where $k_z$ assumes $N_z$ discrete values in the first Brillouin zone \respch{$[-\pi/dx,\,\pi/dx)$}, and the amplitudes $\{u_{n,\sigma}(x,y), v_{n,\sigma}(x,y)\}$ are discretized on an $N_x \times N_y$ mesh.

\begin{figure*}[t]
  \centering
  \includegraphics[width=0.9\textwidth]{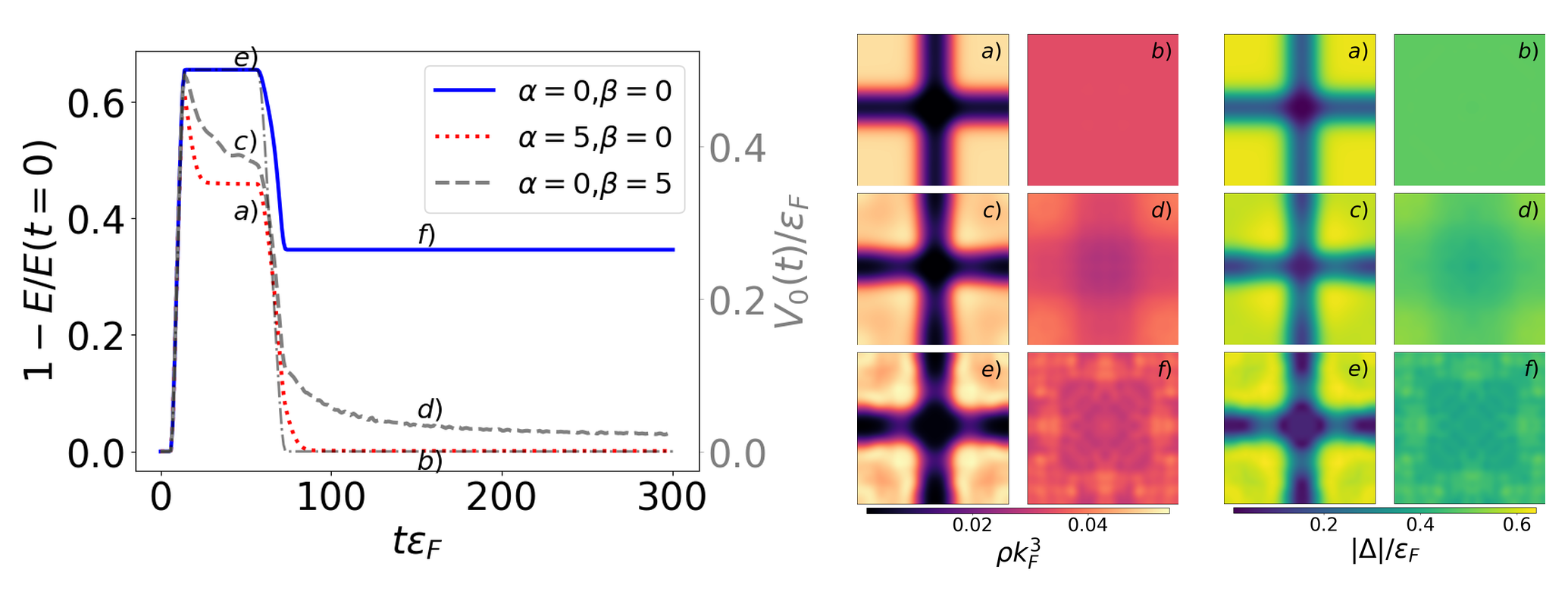} 
  \caption{
    Evolution of the total energy of an initially uniform 2D state in a time-dependent perturbation \cref{eq:testBperturb} with different \respch{cooling} parameters that conserve particle number ($\gamma = 0$).
    The left panel displays the energy dynamics: the solid blue line represents evolution without \respch{cooling}, while the dotted red and dashed gray lines correspond to $(\alpha = 0, \beta = 5)$ and $(\alpha = 5, \beta = 0)$, respectively. The black dotted-dashed line depicts the evolution of the amplitude \respch{$V_0(t)$} of the external potential.
    The right panels display snapshots of the density and pairing fields at the times indicated in the energy plot, with the different \respch{cooling} scenarios separated vertically.
    \label{fig:cooling_2d}
  }
\end{figure*}

\subsection{Cooling a state close to the ground state}\label{ssec:tests1}

The first set of tests examines whether \respch{cooling} dynamics can be used to obtain the ground state of a 1D unitary Fermi gas confined in a harmonic potential.
\respch{Cooling} terms are expected to remove energy from collective degrees of freedom, such as currents (in the case of $\Udiss$).
However, there is no guarantee that in the long-time limit, $t\rightarrow\infty$, the system will relax to the true ground state.
In principle, the dynamics should lead to a stationary state without currents $\vect{j}_\sigma$ and without excitations of the pairing field, while intrinsic quasiparticle excitations may still persist.

In this test, the system is placed in an external harmonic potential,
\begin{equation}
  V_{\textrm{ext}}(x) = \tfrac{1}{2} m \omega^2 x^2,
\end{equation}
with $\omega/\eF(0) = 1.23 \times10^{-2}$, where $\varepsilon_F(0)$ is defined from the
density at the center of the trap.
Because the potential depends only on the $x$ coordinate, all observables (such as density) also depend solely on $x$: the problem is quasi-1D.
The system is solved on a spatial lattice of size $N_x = 256$, and the number of plane waves in perpendicular directions is $N_y\times N_z = 16\times 16$, with a total particle number of $N_\ua = N_\da = 356$.

The initial state is chosen to be close, but not identical, to the ground state.
Specifically, the initial energy exceeds the true ground-state energy by either about $2\%$ or $25\%$.
These configurations are generated by interrupting the iterative ground-state solver before full convergence is reached.
Such situations naturally arise in large-scale simulations where obtaining a fully converged ground state may be prohibitively expensive, and one settles for an approximate solution.
The goal of this test is to determine whether \respch{cooling} dynamics can efficiently remove the residual energy and drive the system toward the true ground state within a time-dependent framework.
Since this system is sufficiently small, the exact ground state is available and serves as a benchmark for comparison.

In \cref{fig:energy_PN_1d} (column a), we present the time evolution of the total energy for four different \respch{cooling} schemes.
When \respch{cooling} is absent ($\alpha=\beta=0$), the total energy remains conserved, and the density exhibits persistent oscillations.
Activating only the mean-field \respch{cooling} channel $\Udiss$ ($\alpha \neq 0$, $\beta = 0$) damps the currents responsible for these oscillations, leading to a gradual decrease of the total energy.
However, even at long times, the system does not relax to the true ground state.
In contrast, \respch{cooling} through the pairing field $\Ddiss$ ($\alpha = 0$, $\beta \neq 0$) proves more effective, in many cases sufficient to cool the system to the ground state.
As expected, the most efficient relaxation occurs when both channels are active ($\alpha \neq 0$, $\beta \neq 0$), which reduces the convergence time.
Columns b) and c) of \cref{fig:energy_PN_1d} show how the density and pairing fields converge after cooling.
These results demonstrate that the proposed method effectively accelerates and improves ground-state preparation, though it does not guarantee convergence to the global energy minimum.

Systematic tests with various \respch{cooling} strengths (see \cref{appendix-B}) show that values of $\alpha \approx 10$ and $\beta \approx 5$ provide a good balance, although optimal values will depend on the exact nature of the system under consideration.
While larger values might initially cool more quickly, they can ultimately saturate.
This is similar to the case of the critical damping coefficient in a harmonic oscillator.
Over-damping delays the cooling of the system by slowing the generation of currents that ultimately allow for the cooling potential to extract energy.
Larger values of $\alpha$ and $\beta$ can also destabilize the numerical integration when \respch{cooling} terms dominate the single-particle Hamiltonian or pairing field.

\subsection{Perturbing a uniform system with a rapid quench}

In the second test, we consider a uniform spin-symmetric unitary Fermi gas with equal particle numbers $N_\ua = N_\da$.
In the absence of an external potential, we can precisely prepare the system in the ground state.
We then apply a localized, time-dependent perturbation of the form
\begin{equation}
  \label{eq:testBperturb}
  V_{\textrm{ext}}(x,y,t) = V_0(t) \Biggl(
  \exp\Bigl(-\frac{x^2}{2\sigma^2}\Bigr)
  +\exp\Bigl(-\frac{y^2}{2\sigma^2}\Bigr)
  \Biggr)\,,
\end{equation}
where $\sigma \approx 2.21 \kF^{-1}$ and the amplitude $V_0(t)$ evolves as follows (left panel in \cref{fig:cooling_2d}):
During the initial interval $t\eF \in (5,15)$ it is ramped up to a value $V_0/\eF = 0.5$, held constant until $t\eF = 55$, and then removed in the interval $t\eF \in (55,75)$.
This quench is strongly nonadiabatic, exciting the system to an energy roughly $35\%$ above the ground state.
Note that, in this case, the potential acts only in the $x$-$y$ plane, so the problem is quasi-2D.
We have used a lattice of size $N_x = N_y = 32$ with $N_z = 16$ plane waves in the $z$ and total particle number $N_\ua + N_\da= 554$. 

\begin{figure*}[t]
  \centering
  \includegraphics[width=\textwidth]{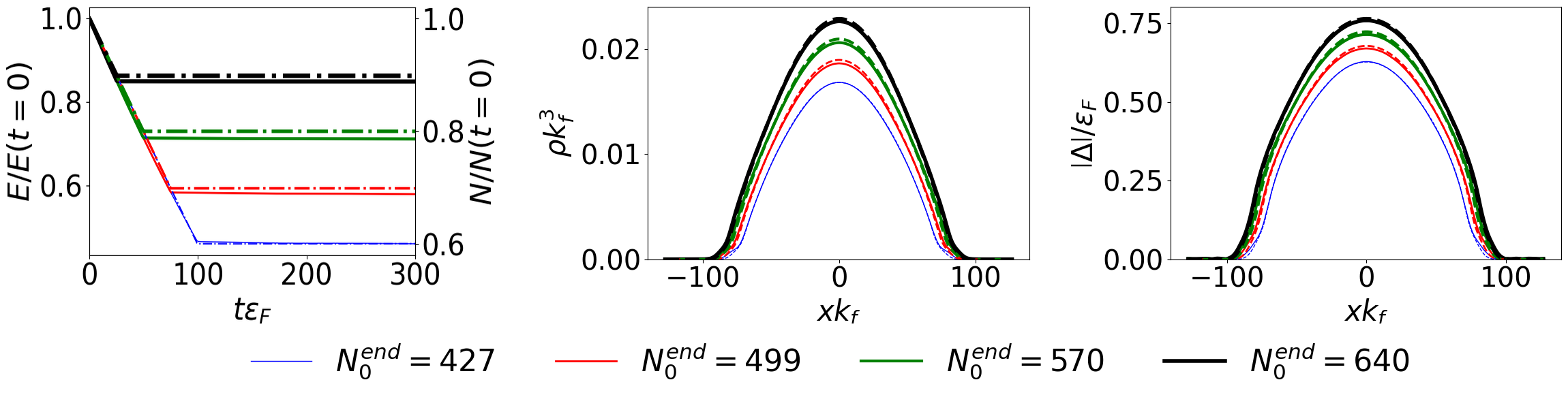}
  \caption{
    Particle-changing time evolution with $\gamma = 0.1$, $\alpha = 10$, and $\beta = 5$, for a system initially containing $N(t=0) = 712$ particles toward different target populations $N_0^{\mathrm{end}} \in \{427, 499, 570, 640\}$.
    We drop $\Nreq $ to the targeted $N_0^{\mathrm{end}}$  at a constant rate  of $\Nreq'(t) = -2.86\eF$.
    Once $N(t) \approx N_0^{\mathrm{end}}$, we resume particle-conserving cooling by setting $\gamma = 0$ and evolve until $t\eF = 300$.
    The left panel shows the time evolution of the total energy (solid lines) and particle number (dash–dotted lines). 
    The center and right panels display the final density and pairing-field profiles at $t\eF = 300$. 
    The ground-state profiles are shown for comparison as dashed lines, with thickness and colors corresponding to the legend below.
    \label{fig:3}}
\end{figure*}

The effect of the perturbation on the total energy is shown in \cref{fig:cooling_2d}.
The solid blue line corresponds to evolution without \respch{cooling}, which plateaus at roughly $35\%$ above the ground-state value for $t\eF > 75$.
Including \respch{cooling} effectively removes this excess energy.
In contrast to the previous test, \respch{cooling} through the mean-field channel (dotted red line) is more effective than cooling with the pairing channel (dashed gray line).
These differences arise because the perturbation directly couples to the density, producing density currents.
Thus, $\Udiss$, which primarily damps currents, provides stronger relaxation.
Panels \respch{a)–f)} show the density and pairing gap at the selected times indicated in the main panel.
Except for panel \respch{b), spatial non-uniform distributions are visible}, illustrating the presence of excitations in the system.
Overall, these results demonstrate that the relative efficiency of $\Udiss$ and $\Ddiss$ depends strongly on the type of excitation introduced, and that combining both mechanisms yields the most effective \respch{cooling} strategy.

\subsection{Changing the particle number}\label{ssec:changpart}

So far, we have considered evolutions in which the particle number is fixed ($\gamma = 0$).
However, as discussed in \cref{ssec:Controlling-particle-number}, there are situations where allowing the particle number to vary during time evolution can be advantageous.
Here, we demonstrate the ability to dynamically control the particle number within the time-dependent \gls{HFB} framework.

We test this implementation using the same harmonic oscillator setup introduced in \cref{ssec:tests1}, but now with non-zero $\gamma>0$ in \cref{eq:tilde-gamma} to guide the evolution toward a desired particle number $\Nreq(t)$.  As shown in \cref{fig:3}, we begin the evolution from the ground state with $N_0 = 712$ particles using the cooling parameters $\alpha$, $\beta$, and $\gamma$. We linearly ramp $\Nreq(t)$ to $N_0^{\text{end}}$ 
\begin{equation}
    \Nreq(t) = \max \bigl[N_0-rt, N_0^{\text{end}}\bigr],
\end{equation}
at a constant rate $r=2.86\eF$.  
Once the desired final particle number is reached $N(t) \approx N_0^{\mathrm{end}}$, we resume particle-conserving cooling by setting $\gamma = 0$ and evolve until $t\eF = 300$.

The rate of change is relatively high; for instance, in the most aggressive scenario, we reduce the particle number to $60\%$ of its initial value within a time interval $t\eF \approx 100$.
Although the instantaneous ground state is not fully reached at intermediate times due to residual excitations, the final density and pairing-field profiles closely reproduce the self-consistent ground-state results.
This agreement arises from the continuous removal of excitation energy through the \respch{cooling} terms $\Udiss$ and $\Ddiss$.

These tests clearly demonstrate that the proposed method enables arbitrary control over the particle-number evolution.
It allows not only for modeling particle losses but also for controlled particle injection.
When the requested particle-number variation is slow compared to the Fermi timescale, the system remains close to its instantaneous ground state throughout the evolution, thereby enabling scans of ground-state properties as a function of density within a time-dependent framework.
This approach leads to a methodology that we refer to as a \respch{density scan}, which will be employed in \cref{sec:neutron-star-crust}.

\section{Applications}
\subsection{Spin-imbalanced unitary Fermi gas}
\glsunset{LO}
\glsunset{FF}
\glsunset{LOFF}
As an application of the \respch{\gls{LQC}} algorithm, we consider spin-imbalanced ultracold Fermi gases at zero temperature for various spin polarizations.
The low-energy states of these systems have been investigated in~\cite{Tuzemen2023}, where it was shown that achieving convergence with static solvers is particularly challenging.
This difficulty arises because the order parameter can develop exotic spatially inhomogeneous superfluid structures that compete with more ordered phases, such as those proposed by Larkin and Ovchinnikov (\gls{LO})~\cite{lo} and independently by Fulde and Ferrell (\gls{FF})~\cite{ff}. 

\begin{figure}[t]
  \centering
  \includegraphics[width=0.73\linewidth,trim={0cm 5cm 0cm 1cm}, clip]{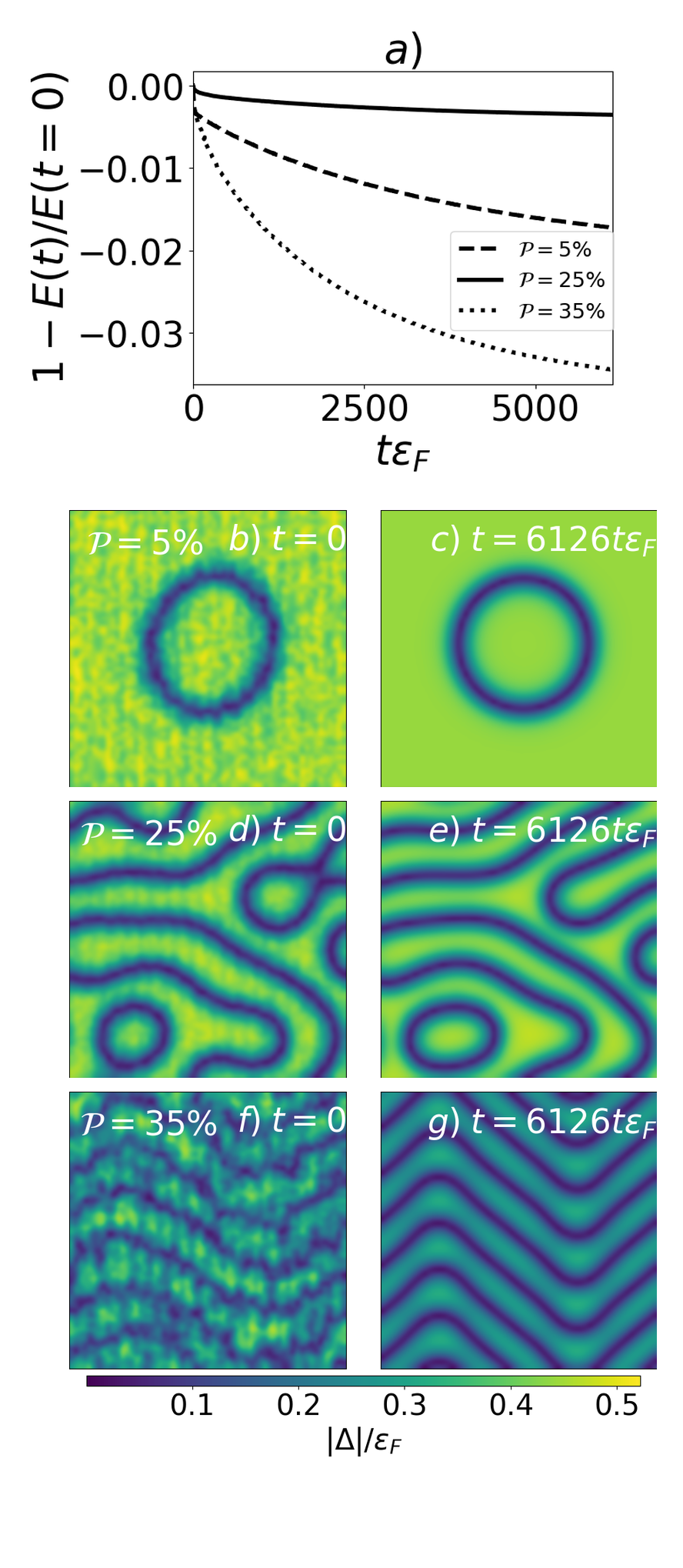} 
  \caption{\respch{a) }Decrease in the total energy of a spin-imbalanced unitary Fermi gas during \respch{cooling} dynamics with parameters $\alpha = 10$ and $\beta = 5$, shown for several spin polarizations $\respch{P}$. 
    \respch{b)–g)} Evolution of the pairing-field magnitude $|\Delta(x, y)|$ under the same conditions. 
    Panels \respch{b), d), and f)} display the initial pairing-field distributions obtained from static ASLDA calculations for different spin polarizations, while panels \respch{c), e), and g)} show the corresponding configurations after \respch{cooling} evolution. 
    The sequence illustrates the transition from localized ferron states at low spin imbalance, through disordered patterns at intermediate imbalance, to spatially ordered, Larkin–Ovchinnikov-like (LO-like) phases at high imbalance.
    \label{fig:spinimb-run}}
\end{figure}
Here, we test whether the qualitative results reported in~\cite{Tuzemen2023} persist when the ground state is obtained using the \respch{cooling} strategy described in \cref{ssec:tests1}. 
\textcite{Tuzemen2023} found that, at low spin imbalances
\begin{equation}
  \respch{P} = \frac{\abs{N_\ua - N_\da}}{N_\ua + N_\da} \lesssim 5\%\,
\end{equation}
the ground state is characterized by localized excitations such as ferrons~\cite{PhysRevA.100.033613} or ring solitons~\cite{PhysRevResearch.2.043282}.
These are spatially localized structures with radially modulated spin density, leading to the formation of nodal rings.

For moderate spin imbalances $\respch{P} \lesssim 30\%$, the system self-organizes into a disordered pattern reminiscent of liquid-crystal-like phases.
At larger imbalances $\respch{P} \gtrsim 30\%$, periodic structures emerge, and the system approaches a long-predicted~\cite{PhysRevLett.101.215301} \gls{LO}-like state.

To carry out this test, we first obtain an approximate solution using the \gls{ASLDA} following the protocol outlined in~\cite{Tuzemen2023}.
As the static solver did not achieve full convergence to the desired accuracy, the initial state is slightly excited about the ground state, 
Calculations are performed on a spatial lattice of size $N_x \times N_y = 60 \times 60$, with $N_z = 16$ plane-wave modes along the $z$ direction, so the problem is quasi-2D with the density and pairing fields being independent of $z$.
Representative initial states for different spin imbalances obtained from partially-converged static calculations are shown in panels \respch{b), d), and f) of \cref{fig:spinimb-run}.
The corresponding energy decrease from the \respch{\gls{LQC}} algorithm is shown in panel a)}. 

The nodal lines visible in the pairing field $\Delta(x,y)$ (dark blue regions where $\Delta$ vanishes) correspond to locations where the majority-spin density dominates, i.e., $\rho_{\ua\ua}(\vect{r}_{\textrm{nodal}}) \gg \rho_{\da\da}(\vect{r}_{\textrm{nodal}})$.
Conversely, in regions where $|\Delta|$ is large, the spin densities are nearly equal, $\rho_{\ua\ua}(\vect{r}) \approx \rho_{\da\da}(\vect{r})$.
The presence of short-wavelength fluctuations in the initial pairing-field maps indicates that these configurations are not fully converged static solutions and still carry residual excitation energy.
This excess energy allows these systems to explore the potential energy landscape during evolution, potentially relaxing into a state with qualitatively different properties.

For low-spin imbalances $\respch{P} \approx 5\%$, \respch{\gls{LQC}} drives the system toward a configuration containing a single, spherically shaped ferron \cref{fig:spinimb-run}(b–c).
This behavior is consistent with earlier findings~\cite{Tuzemen2023,PhysRevResearch.2.043282} which showed that increasing spin imbalance does not trigger an immediate transition from the uniform ground state to an \gls{LO} or \gls{FF} phase.
Instead, intermediate states featuring ferrons appear.
Because ferrons are not topologically protected, they can, in principle, transform smoothly into other nonuniform structures or dissipate entirely, leading to a uniform polarized state.
However, such relaxation to uniformity is not observed in our simulations.
Even for long evolution times, the total energy does not settle to a well-defined value.
This is similar to static-calculation results, where thousands of self-consistent iterations were empirically required to approximate the ground state (or local minimum).

As the spin imbalance increases, e.g., to $\respch{P} \approx 25\%$, the static solver typically yields disordered configurations.
To determine whether such disorder is an intrinsic property of the ground state or merely a computational artifact, we apply the \respch{\gls{LQC}} algorithm to see if the system relaxes toward a more ordered configuration.
Although \respch{cooling} removes excess excitation energy and smooths the nodal-line patterns, the overall disordered nature of the state persists \cref{fig:spinimb-run}(d–e).
The results obtained across a range of spin polarizations indicate that disordered states are indeed characteristic of the spin-imbalanced unitary Fermi gas in this intermediate polarization regime.

Finally, for larger imbalances $\respch{P} \gtrsim 35\%$, both the static solver and the subsequent \respch{\gls{LQC}} evolution lead to states that exhibit spatial order in their lowest-energy configuration.
This is illustrated in \cref{fig:spinimb-run}(f–g), where an initially noisy configuration evolves into an ordered structure, consistent with the expected \gls{LO}-like pattern.

In summary, we have tested if the qualitative features of low-energy states in a spin-imbalanced unitary Fermi gas persist when the state is cooled with \respch{\gls{LQC}}.
Consistent with~\cite{Tuzemen2023}, we find: at low spin imbalance, the ground state forms ferron-like structures; at moderate imbalance, disordered states dominate; and at high imbalance, the system transitions into a spatially ordered \gls{LO}-like phase.

\begin{figure}[t]
  \centering
  \includegraphics[width=0.95\linewidth]{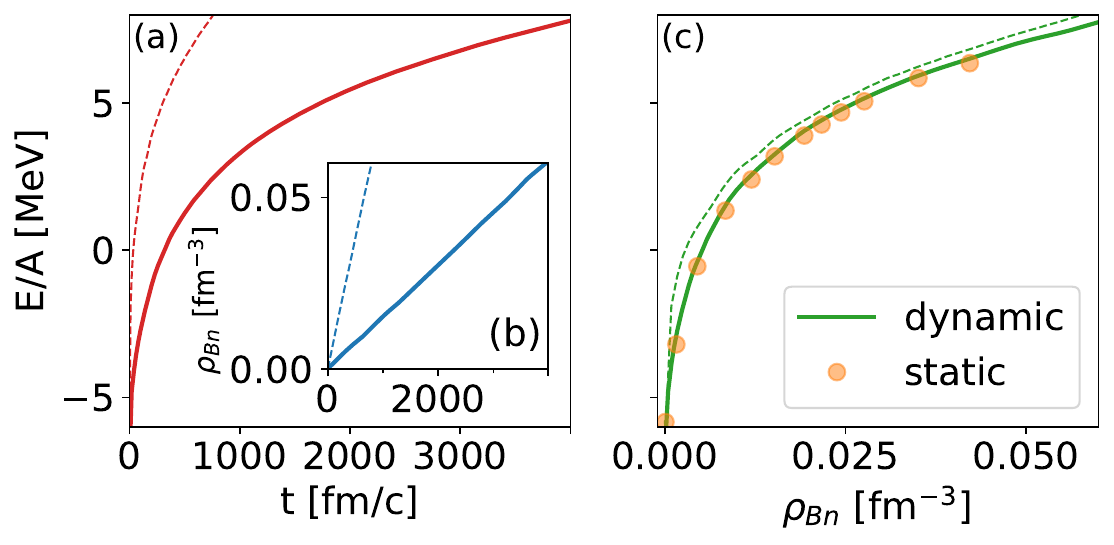} 
  \caption{Time-dependent \respch{density scan} of a Zirconium nucleus ($Z=40$) in a neutron background with increasing density.
    (a) Total energy as a function of evolution time. 
    (b) Evolution of the neutron background density $\rho_{Bn}$. 
    (c) Total energy per particle as a function of $\rho_{Bn}$ (solid curve), compared with independent static calculations~\cite{pecak2024WBSkMeff} (circles).
    The dashed line in all frames corresponds to a run where the rate of neutron injection into the system was increased by a factor of five.
    This demonstrates that a single \respch{density scan} can accurately reproduce multiple static simulations, provided the background density changes sufficiently slowly.
    \label{fig:ns-scan-1}}
\end{figure}

\subsection{Neutron star crust}
\label{sec:neutron-star-crust}
We now consider nuclear matter, whose density is roughly 30 orders of magnitude larger than that of ultracold gases.
Bulk nuclear matter exists naturally in neutron stars, the remnants of cataclysmic supernova collapses.
We focus, in particular, on the outer crust of neutron stars, where a Coulomb crystal of charged nuclei \respch{is embedded in} a sea of superfluid neutrons.
Due to the large neutron-neutron scattering length, this dilute neutron superfluid is closely related to the unitary Fermi gas, despite the huge difference in scales.
A quantitative analysis, however, requires a more careful accounting of nuclear interactions.
To this end, we use the W-BSk Toolkit~\cite{pecak2024WBSkMeff}, which is optimized for such systems and based on the BSk31 density functional of the Brussels–Montreal type~\cite{goriely2009skyrme, goriely2016further}.
Since \cref{eq:alphabeta} is very general, there is considerable flexibility in choosing $\alpha_{\mu\nu}$ and $\beta_{\mu\nu}$.
To check the robustness of our method, we follow a simple prescription that assumes a very basic form of the energy density functional \cref{eq:hss}.
In principle, one can derive formulas for more general forms of the energy functional (see \cref{appendix-C}).

\label{sec:density-scan}
\begin{figure*}[t]
  \centering
  \includegraphics[width=0.95\linewidth]{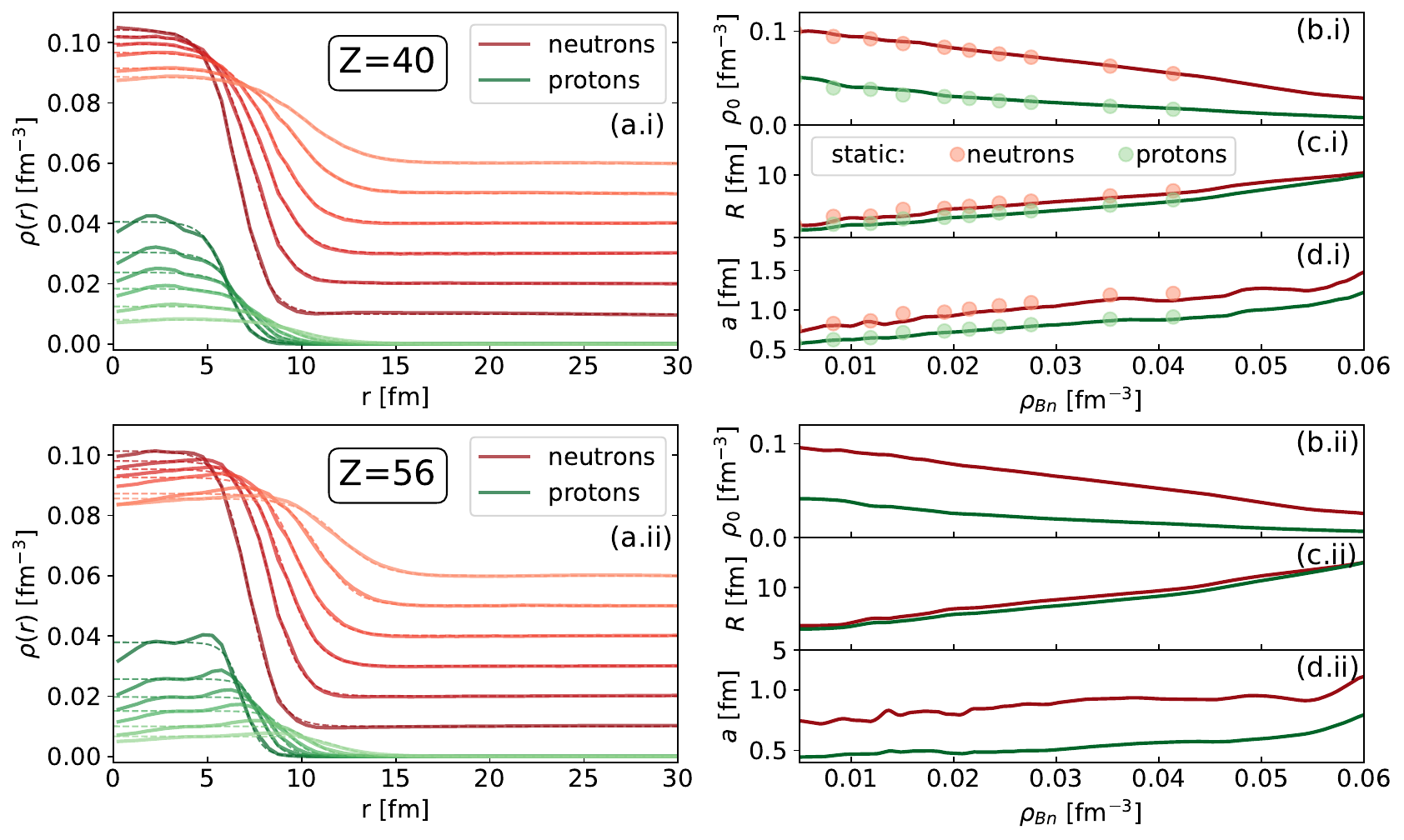} 
  \caption{
    Structural properties of protonic clusters in the neutron-star inner crust obtained from time-dependent \respch{density scan}s. 
    The top row (i) is Zirconium $Z=40$ while the bottom row (ii) is Barium $Z=56$. 
    (a) Radial density profiles for neutrons and protons at selected neutron background densities $\rho_{Bn}$. 
    Dashed lines represent the best-fit profiles~\cref{eq:densprofile}.
    Fading colors indicate a correspondence to longer simulation times and larger background neutron density.
    (b–d) Extracted parameters of the fitted profiles: (b) central density $\rho_{0q}$, (c) nuclear radius $R_q$, and (d) surface thickness $a_q$ as functions of $\rho_{Bn}$ for both neutrons and protons. 
    The results demonstrate a gradual decrease in central density and a simultaneous increase in surface diffuseness and nuclear radius with rising neutron density. 
    For $Z=40$, parameters extracted from static calculations of~\cite{pecak2024WBSkMeff} are shown as dots.
    \label{fig:ns-scan-2}}
\end{figure*}

\subsubsection{Properties of nuclei immersed in neutron matter}
For crust densities $\rho \lesssim \qty{0.04}{fm^{-3}}$, where $\rho$ denotes the total nuclear density (neutrons and protons), protons and neutrons are expected to bind into quasi-spherical nuclei immersed in a dilute neutron superfluid.
Calculating properties such as sizes or surface smoothness remains an active research area in neutron-star physics~\cite{Negele1973,PhysRevC.85.035801,PhysRevC.94.065801,PhysRevC.87.035803,PhysRevC.110.045808,pecak2024WBSkMeff}.
The standard approach involves analyzing each background density separately and, for each value, determining the ground state of the nucleus.
Here, we adopt an alternative strategy: We generate a single static solution and then evolve it dynamically, gradually increasing the number of neutrons in the system with \respch{\gls{LQC}}.
In this way, within a single time-dependent simulation, we can scan the properties of the ground state of the nucleus as a function of the changing background neutron density.

\Cref{fig:ns-scan-1} shows an example of such a simulation.
We begin with a solution obtained in a cubic box of size $32^3$ with a lattice spacing of $\d{x} = \qty{1.25}{fm}$.
We start with $Z = 40$ protons and $N_0 = 116$ neutrons.
Since the number of neutrons already exceeds the neutron drip line, the initial configuration corresponds to a Zirconium nucleus immersed in a dilute neutron gas.
We then evolve this system in time with \respch{\gls{LQC}} ($\alpha = 10$, $\beta = 5$) and include a particle-changing term $\gamma=10$ to gradually increase the number of neutrons with
\begin{align}
  \Nreq(t) = N_0 + r t\
\end{align}
in \cref{eq:tilde-gamma}.
The number of protons remains fixed.

With a sufficiently slow scan, $r = \qty{1}{$c$/fm}$, the \respch{\gls{LQC}} algorithm keeps the system close to the ground state, allowing us to deduce the properties of the Zirconium nucleus over a wide range of background neutron densities $\rho_{Bn}(t)$ (evaluated far from the Zirconium nucleus).
The resulting energies per particle are compared with many independent static calculations from~\cite{pecak2024WBSkMeff}, showing remarkable agreement between the two approaches.
The key difference lies in computational efficiency: while the static approach requires finding a separate \ce{^{40}Zr_{$N$}} solution for each neutron number $N$, the time-dependent approach yields an entire spectrum of solutions with a single \respch{density scan}.
We have also repeated the calculation with $r=\qty{5}{$c$/fm}$, changing the background density five times faster.
With this faster scan, the \respch{\gls{LQC}} algorithm cannot track the ground state as well, and the system remains excited, resulting in overestimated energies.

In \cref{fig:ns-scan-2} we use the same method to study the properties of two nuclei: Zirconium ($Z=40$) and Barium ($Z=56$) immersed in a dilute neutron background.
Panel (a) shows how the neutron and proton density profiles change as we scan through increasing values of the background density $\rho_{Bn}(t)$.
With reasonable accuracy (better for neutrons than for protons), these profiles can be parameterized as
\begin{gather}
  \label{eq:densprofile}
  \rho_\sigma(r) = \frac{\rho_{0\sigma}}{1+\exp\left( \frac{r-R_\sigma}{a_\sigma} \right)} + \rho_{B\sigma},
\end{gather}
where $\sigma \in \{n, p\}$ is the isospin species, $R_\sigma$ is the nuclear radius, $a_\sigma$ is the surface thickness, and $\rho_{0\sigma}$ and $\rho_{B\sigma}$ are the densities inside and outside the nucleus (with $\rho_{Bp} = 0$ for protons).

The extracted parameters are presented in panels (b)–(d).
As the neutron background density increases, the central densities $\rho_{0n}$ and $\rho_{0p}$ decrease.
Note that in the neutron-star crust, $\rho_{0n} + \rho_{0p} < \rho_0$, where $\rho_0 \approx \qty{0.16}{fm^{-3}}$ is the nuclear saturation density.
The surface thickness increases with neutron density and is consistently larger for neutrons than for protons.
In contrast, the radii of the neutron and proton distributions remain comparable $R_n \approx R_p$ and increase with the background neutron density.
For $Z = 40$ we also compared the extracted shape parameters with those obtained using the static approach of~\cite{pecak2024WBSkMeff}, again finding good agreement between the two methods.

While the time-dependent method enables computationally efficient \respch{density scans} of protonic clusters in the neutron-star crust, it does not allow for the study of transitions between different structural configurations.
In particular, for $\rho_{Bn} \gtrsim \qty{0.05}{fm^{-3}}$, the system is expected to undergo transitions to other configurations like rod-like or planar proton distributions (so-called \respch{nuclear pasta} phases).
Typically, these configurations are separated by energy barriers.
Since our calculations dampen fluctuations during the evolution, the system remains trapped near a local minimum corresponding to the cluster-like nuclear configuration.
In the next section, we will demonstrate how \respch{\gls{LQC}} can be used to obtain states of different structures, relevant for modeling the neutron-star crust at higher densities.

\subsubsection{Nuclear Pasta}
Obtaining the ground state using standard methods requires numerous diagonalizations, which are computationally expensive.
The cost of a single diagonalization scales with the cube of the domain size, $N^3$, and quickly dominates over the cost of evolution in time-dependent frameworks.
This challenge becomes particularly severe at higher densities in the crust of neutron stars, where nuclei are expected to deform into so-called \respch{nuclear pasta} phases~\cite{caplan2017pasta}.
In these layers, the repulsive Coulomb energy becomes comparable to the attractive nuclear energy: these contributions almost cancel, requiring highly accurate numerical calculations to enable a meaningful comparison between the energies of these phases.
Moreover, these systems contain numerous local energy minima, which slow the convergence toward the true ground state. 
The \respch{\gls{LQC}} algorithm gives a practical tool for studying nuclear pasta configurations by gradually removing excess energy.
In this case, the extension to include \respch{cooling} in the pairing field $\beta \neq 0$ is particularly important, since the superfluidity of both protons and neutrons contributes at high densities.

\begin{figure}[t]
  \centering
  \includegraphics[width=.99\linewidth]{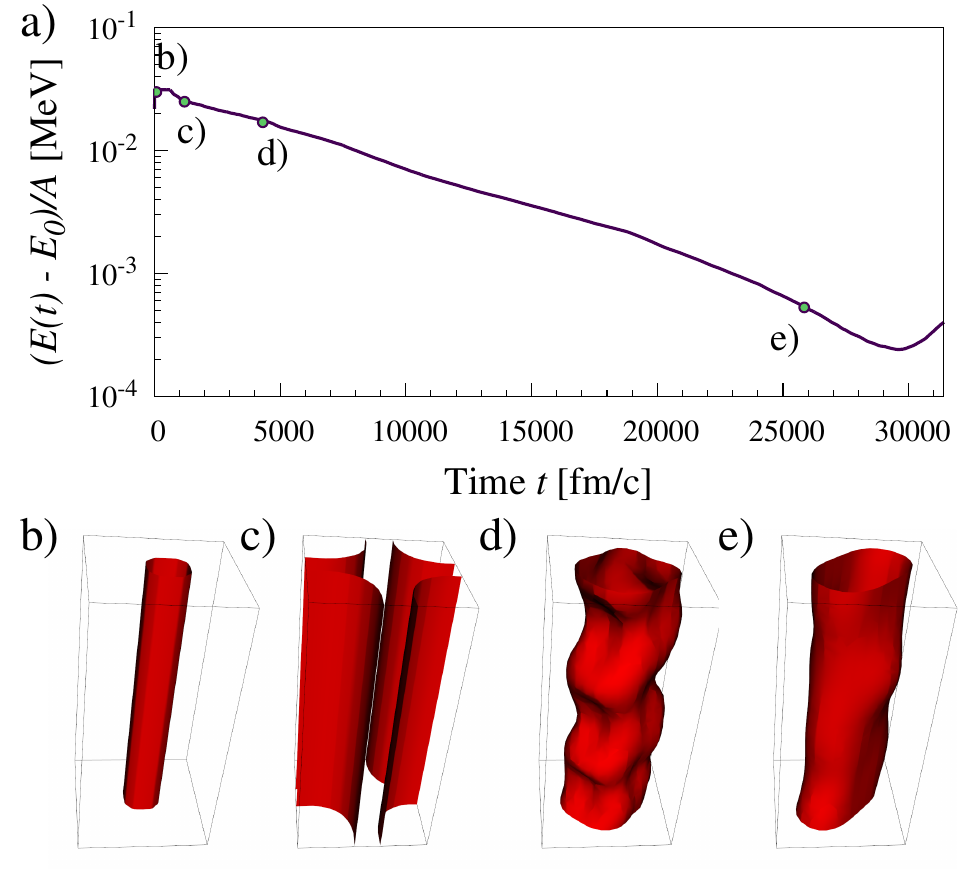}
  \caption{\label{fig:ns-single-rod}
    a) Evolution of the energy difference with the approximate ground state energy $E_0=\qty{6.035}{MeV}$) while evolving with \respch{\gls{LQC}}.
    The calculation starts from a uniform system, where a cylindrical potential is briefly applied to break the symmetry.
    Near $t \approx \qty{29000}{fm/$c$}$, numerical errors accumulate, and the method can no longer be continued.
    Panels b) through e) show isosurfaces of the proton density $\rho_p = \qty{0.0029}{fm^{-3}}$ at times $t = \qtylist{81;1200;4312;25843}{fm/$c$}$, respectively.
  }
\end{figure}

In \cref{fig:ns-single-rod} we show how a rod-like structure (i.e., the spaghetti phase) develops using \respch{\gls{LQC}}.
We start with a uniform nuclear matter solution, with the $Z=60$ protons and $N=1796$ neutrons, which will be conserved in the subsequent time evolution.
Next, an external potential of cylindrical shape is applied for a short time interval (\qty{80}{fm/$c$}) to break the symmetry of the system.
This perturbation determines the generic shape emerging from the evolution as shown in \cref{fig:ns-single-rod}(b).
After the potential is removed, the system evolves toward the low-energy state configuration.
The energy differences (per nucleon) are on the order of tens of \unit{keV}; however, even such small changes drive the system through intermediate shapes of various geometry, ultimately reaching the final state shown in \cref{fig:ns-single-rod}(e). 

\begin{figure}[ht]
  \centering
  \includegraphics[width=.99\linewidth]{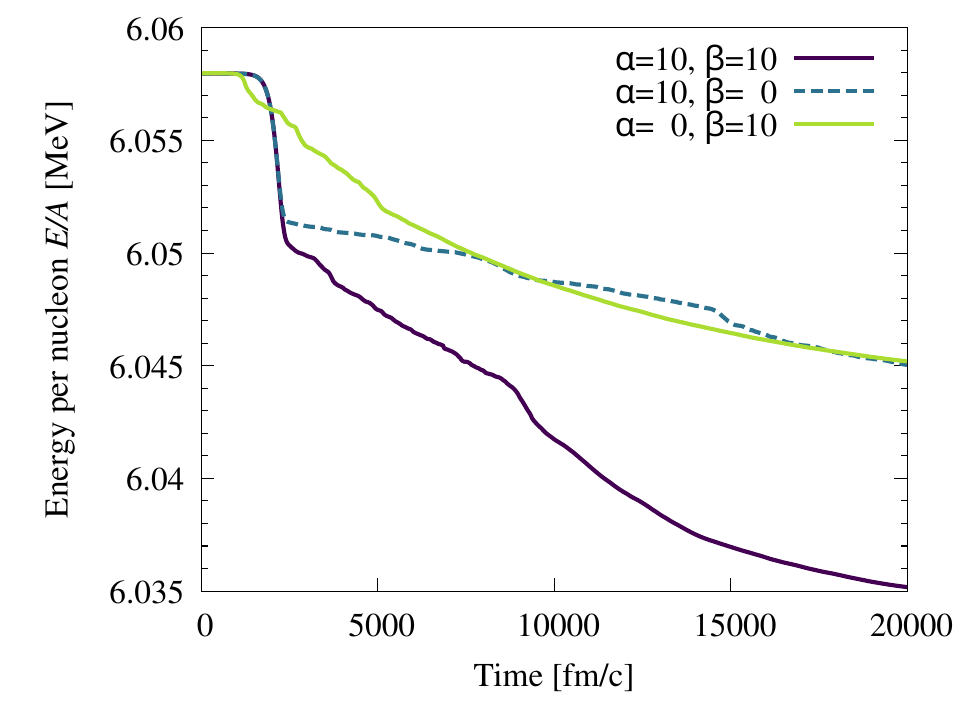}
  \caption{Effect of different parameter values $\alpha$ and $\beta$ on the rate of energy loss.
    The \respch{\gls{LQC}} algorithm is most efficient when both parameters are nonzero, similar to the case of ultracold gases shown in \cref{fig:cooling_2d}.
    \label{fig:cooling_ns}}
\end{figure}
As for ultracold gases (see \cref{fig:cooling_2d}), we compare different combinations of $\alpha$ and $\beta$ to achieve optimal energy removal.
As shown in \cref{fig:cooling_ns}, we find that, although the optimal choice remains system-dependent, it is again a good practice to use cooling in both density and pairing channels.

\begin{figure}[ht]
  \centering
  \includegraphics[width=.99\linewidth]{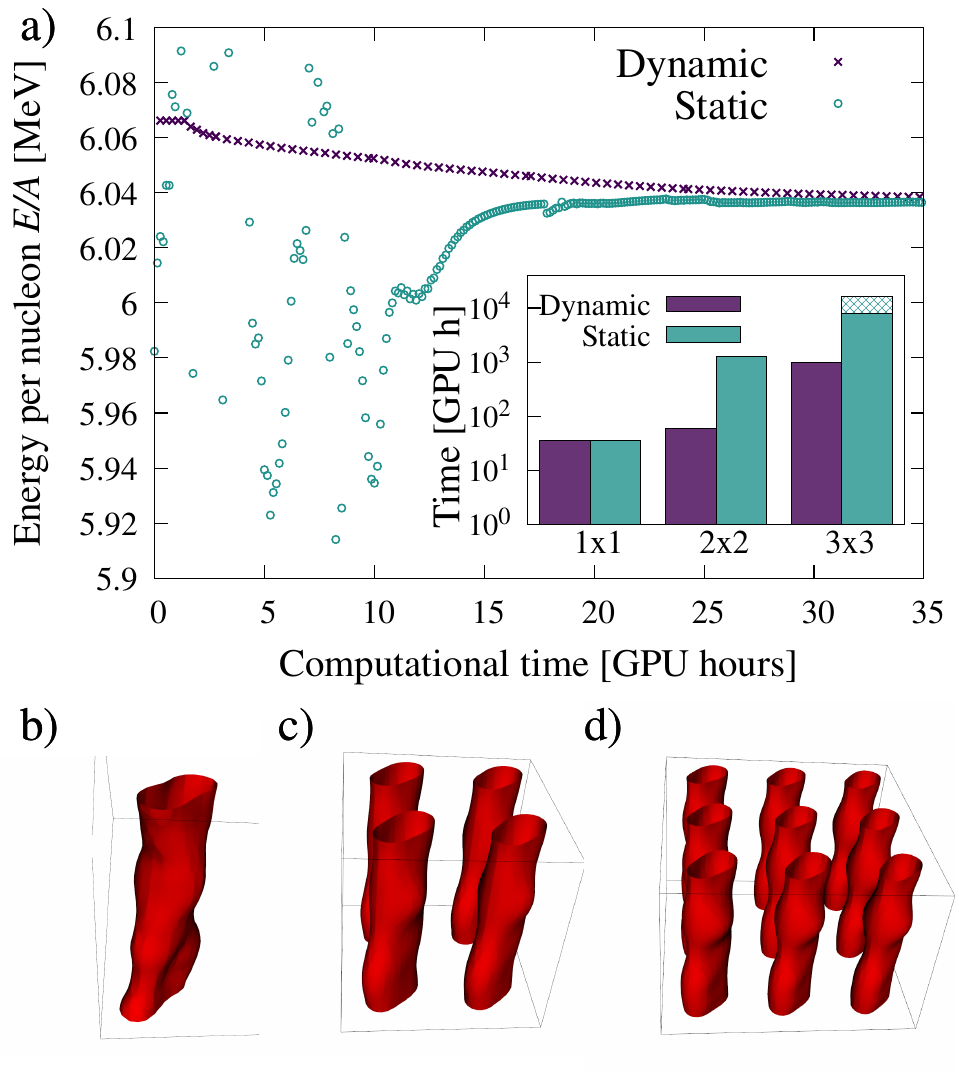}
  \caption{\label{fig:ns-grid}
    a) Comparison of static versus dynamic approaches to obtain a well-converged state of nuclear spaghetti (see text for details).
    The inset shows the estimated computational time to obtain a converged solution with the dynamic (violet) and static (teal) approaches. 
    Bottom panels show isosurface of proton density $\rho_p=\qty{0.005}{fm^{-3}}$ for: b) a single rod; c) $2\times 2$ grid; d) $3\times3$ grid.
    System sizes and computational costs increase by an order of magnitude in every case.
  }
\end{figure}

In \cref{fig:ns-grid}(a), we compare the computational cost to obtain a nuclear rod (spaghetti phase) using the dynamical \respch{\gls{LQC}} approach and the traditional static method based on repeated diagonalizations.
Both lead to the same physical solution -- a rod-like configuration -- after starting from a uniform system at a given density and then breaking the symmetry with an external potential.
When searching for a solution comprising a single rod, i.e., considering relatively small simulation volumes, the computational cost is comparable for the two methods, amounting to approximately 30 GPU-hours on AMD MI250x \glspl{GPU} at the LUMI supercomputer.
The simulations use a grid of $20 \times 20 \times 40$ points with a lattice spacing of \qty{1.2}{fm}.
At this scale, both methods have similar computational costs.
The main difference, illustrated in \cref{fig:ns-grid}(a), is that the dynamical approach evolves continuously in time, whereas the static approach proceeds through discrete diagonalization steps. In the dynamical scheme, the particle number can be conserved and fixed by the initial conditions. In contrast, the static method may traverse intermediate states with different particle numbers, since the chemical potential depends sensitively on the current configuration and must be iteratively adjusted to reach the final state with the desired particle content. As a consequence, in the dynamical evolution, the energy approaches the ground-state value from above, while in the static approach, only the final converged state has the correct particle number and the corresponding ground-state energy.

In a large volume, the spaghetti phase consists of a lattice of rod-like structures.
To model this more realistic configuration, we replicate a single rod solution multiple times to construct the ground state of a periodic grid; see \cref{fig:ns-grid}(b) for a $2\times2$ grid and \cref{fig:ns-grid}(c) for a $3\times3$ grid.
These panels show the proton-density isosurfaces within the considered region of the inner crust.
By replicating the system, we obtain a reasonable initial configuration for simulations in a larger computational domain.
However, duplicating a single solution introduces discontinuities at the matching interfaces, which hinders convergence in the diagonalization method and significantly increases the computational time required.
We compare the convergence times for each system in the inset of \cref{fig:ns-grid}(a).
For the single rod, the resource requirements are similar for both methods.
For the $2\times2$ grid, the static solution converges toward the dynamical one after roughly 1200 GPU hours, about an order of magnitude longer than the dynamical evolution.

In the last example -- the largest system we consider here -- the dynamical solution saturates relatively quickly.
In contrast, the static method exhibits discontinuous behavior and only begins to converge continuously after approximately 8000 GPU hours (solid bar in the inset).
We estimate that full convergence would require at least \num{16000} GPU hours, as shown in the inset, emphasizing the significantly higher computational cost of the static method compared to the dynamical \respch{\gls{LQC}} approach.

\Cref{fig:ns-grid}(b--c) correspond to systems of sizes $40\times 40\times 40$ and $60\times 60\times 40$, respectively.
Larger systems require more computational nodes and thus more communication, causing the computational cost to scale accordingly, most significantly impacting the static approach, which requires all-to-all communication.
Thus, we find that the \respch{\gls{LQC}} algorithm can significantly reduce resource usage for large-scale problems.
We believe that this approach could be a very useful tool for generating initial states with multiple rods, which may serve as a starting point for dynamically studying the stability of nuclear pasta structures.

\glsreset{LQC}
\glsreset{TDDFT}
\glsreset{TDSLDA}
\section{Conclusion}
We have developed a new scheme for cooling dynamical simulations called \respch{\gls{LQC}} that can be efficiently applied to large Fermi systems, including superfluids.
The proposed approach is computationally relatively inexpensive \respch{and remains compatible with density functional theory frameworks, such as the \gls{TDSLDA} and its extensions, enabling direct implementation in large-scale solvers used for nuclear~\cite{Jin2021,pecak2024WBSkMeff} and ultracold-atom systems~\cite{Bulgac:2007, Bulgac2019}.}

Within this framework, we have demonstrated that energy can be dissipated through multiple channels.
Besides the previously studied current-damping mechanism~\cite{bulgac2013f,huangphd,Unitary_evolution_Bulgac}, our results reveal that \respch{cooling} through the pairing channel can play a significant role.
In numerous cases, particularly for strongly paired systems, we find that damping of the pairing field is more effective in driving the system toward equilibrium than damping only currents.
This opens a new path for controlling the relaxation of superfluid systems and suggests that the pairing channel could serve as a natural route for modeling particle–particle scattering processes that are otherwise absent in mean-field approaches~\cite{PhysRevC.105.L021601}.

We have also formulated a time-dependent evolution scheme that allows for variation in particle number.
This enables simulations in which the particle number can be guided during evolution \respch{in accordance with prescribed constraints.}
This feature enables us to perform a \respch{density scan}, in which the system is gradually driven through different average densities, thereby mapping out multiple equilibrium configurations with a single time-dependent run.

The generality of the method has been demonstrated through its application to two physically distinct and computationally demanding systems: (i) spin-imbalanced unitary Fermi gases, and (ii) nuclear matter in the neutron star crust.
For the former, we confirmed the qualitative findings of \textcite{Tuzemen2023}, reproducing the formation of inhomogeneous superfluid phases and the rearrangement of the pairing field under spin polarization.
For the latter, we showed that the \respch{\gls{LQC}} algorithm enables efficient preparation of complex inhomogeneous configurations, such as \respch{nuclear pasta} states, and provides a practical tool for performing systematic density scans in this regime.
These capabilities pave the way toward future simulations of vortex dynamics and quantum turbulence in the pasta phase, an area that remains largely unexplored~\cite{RevModPhys.89.041002}.
This approach opens the possibility of comparing large volumes of pasta phases obtained with the \gls{HFB} method to those generated using other approaches, such as liquid-drop, semiclassical, or \gls{HF} models~\cite{schuetrumpf2019survey, shchechilin2025filling}.
Moreover, we demonstrate the readiness to perform time evolution of demanding pasta phase within the full TDHFB method, including the effects of superfluidity, for box sizes and particle numbers comparable to those accessible with classical molecular dynamics methods~\cite{horowitz2005dynamical}.

Here, we have only considered dissipation \respch{via our cooling potential}, leaving stochastic noise generation for future work.
Together, these could lead to a self-consistent framework for studying the interplay of fluctuations and dissipation at finite temperatures.
Such an extension would open the door to modeling thermalization and decoherence in strongly interacting Fermi systems~\cite{Haake2018, PhysRevResearch.6.L042003}.
Of particular interest is the inclusion of stochasticity in the pairing channel, which could effectively mimic particle–particle collisions and bridge the gap between mean-field and kinetic approaches~\cite{PhysRevC.105.L021601}.
We expect that such an approach will enable fully microscopic simulations of finite-temperature relaxation, turbulence decay, and vortex–impurity interactions in both ultracold atomic gases and neutron-star matter.

\section{Acknowledgements}
This work was financially supported by the (Polish) National Science Center Grants No.\ 2022/45/B/ST2/00358 (J.E.A.A., G.W.) and No.\ 2024/55/D/ST2/01516 (D.P.) 
We acknowledge the Polish high-performance computing infrastructure PLGrid for awarding this project access to the LUMI supercomputer, owned by the EuroHPC Joint Undertaking, hosted by CSC (Finland) and the LUMI consortium through PLL/2024/07/017603.
M.M.F.\ acknowledges \supportfromNSFgrant[PHY]{2309322}.
We thank the Institute for Nuclear Theory at the University of Washington for its kind hospitality and stimulating research environment.
This research was supported in part by the INT's U.S. Department of Energy grant No.\ DE-FG02- 00ER41132.
We especially thank Aurel Bulgac and Piotr Magierski for many useful discussions.
\paragraph*{Author contributions:}
The method was developed by M.M.F. and G.W. Its implementation was carried out by J.E.A.A. and G.W., and testing was performed by J.E.A.A. Applications to ultracold gases were conducted by J.E.A.A., while applications to neutron-star systems were performed by D.P. and G.W. All authors contributed to the interpretation of the results and to the writing of the manuscript.

\section{Data availability}
The data that support the findings of this article are openly available~\cite{zenodo}.

\appendix
\section{General form for Cooling}
\label{appendix-A}
Under a cooling potential $\UU$, we have \cref{eq:ENdot}
\begin{align}
  \dot{E} &= -\Tr\bigl(\I[\RR, \tfrac{1}{2}\HH]\UU\bigr)\,,
  & \dot{N} &= -\Tr\bigl(\I[\RR, \NN]\UU\bigr)\,.
\end{align}
We can express this using the Frobenius inner product
\begin{subequations}
  \begin{gather}
    \braket{\mathscr{A}|\mathscr{B}} \equiv \braket{\mathscr{A},\mathscr{B}}_{F}
                                     = \Tr(\mathscr{A}^\dagger\mathscr{B})\,:
  \end{gather}    
  \begin{align}
    \dot{E} &= -\braket{d{E}|\UU}\,,
    & \ket{d{E}} &\equiv \I\ket{[\RR, \tfrac{1}{2}\HH]}\,,\\
    \dot{N} &= -\braket{d{N}|\UU}\,,
    & \ket{d{N}} &\equiv \I\ket{[\RR, \NN]}\,.
  \end{align}
\end{subequations}
The most general form of the cooling potential is a vector $\ket{\UU}$ in the half-space with positive overlap $\braket{d{E}|\UU} > 0$.
We can simultaneously adjust the particle number if we can ensure either $\braket{d{N}|\UU} > 0$ (cooling will lower the particle number), or $\braket{d{N}|\UU} < 0$ (cooling will increase the particle number).

In \cref{eq:alphabeta}, to make the cooling potential local, we consider a restricted subspace where all components of $\ket{\UU}$ have the same sign in some basis -- the position basis for standard locality.
This amounts to considering evolution by a set of ``local'' vectors in the orthant containing $\ket{dE}$.
This is the set of vectors $\ket{\UU}$ where only the diagonal components of the matrix $\UU$ in position space are non-zero.

\begin{figure*}[t]
  \centering
  \includegraphics[width=.9\linewidth]{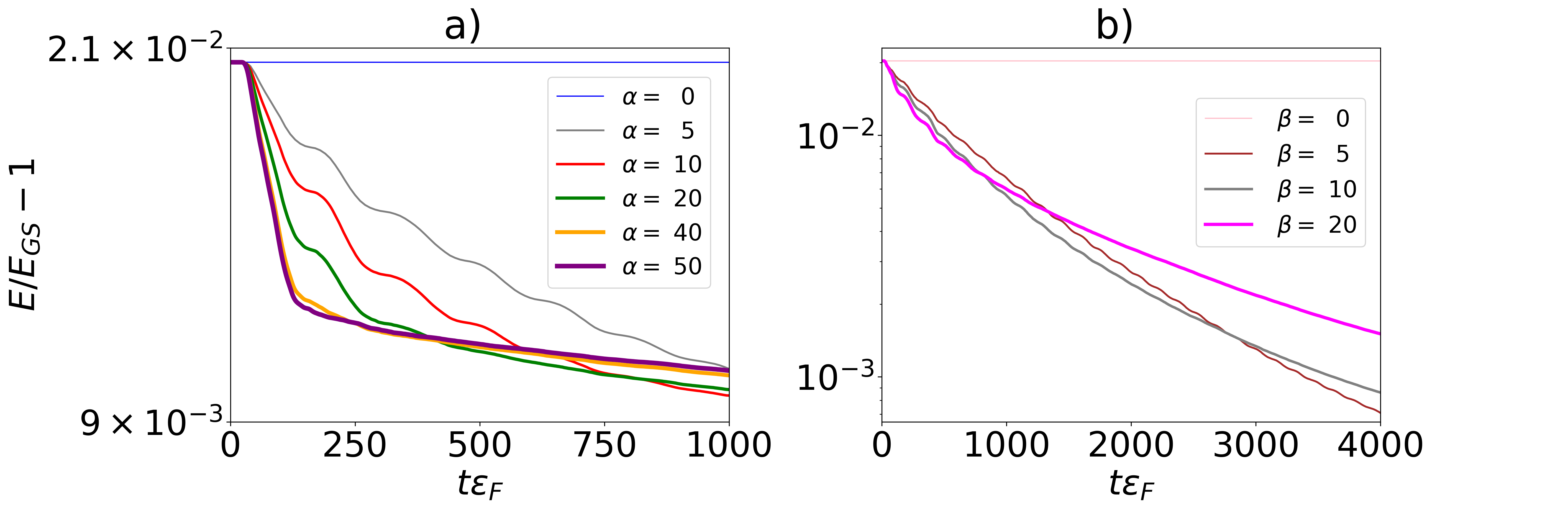}
  \caption{
    Comparison of different \respch{cooling strengths for a) the mean field ($\beta=0$) and b)} the pairing field ($\alpha=0$). 
    The simulated system is the same as that discussed in \cref{fig:energy_PN_1d}.
    \label{fig:OptimalAlphaBeta}
  }
\end{figure*}
 
\section{Optimization of \texorpdfstring{$\alpha, \beta$}{alpha, beta}}
\label{appendix-B}

Here, we demonstrate that there exists an optimal range for the parameters $\alpha$ and $\beta$, which control the strength of the \respch{cooling} terms. 
As shown in \cref{eq:cooling_alpha,eq:tilde-beta}, these parameters are dimensionless. 
We consider the same system as in \cref{ssec:tests1}, but now we systematically investigate the efficiency of \respch{cooling} as the values of $\alpha$ and $\beta$ are varied. 
The results are presented in \cref{fig:OptimalAlphaBeta}.

As shown in panel (a) of \cref{fig:OptimalAlphaBeta}, when energy is dissipated through the mean field, large values of $\alpha$ (e.g., $\alpha \simeq 40$) lead to an overdamped regime: a substantial amount of energy is removed at the beginning of the evolution, but the damping of the current contributions becomes too strong, causing the cooling rate to stagnate. 
Lower values of $\alpha$ (e.g., $\alpha = 5$) are also suboptimal, as they result in slower, although steadier, cooling. 

A similar trend is observed for the pairing field, as shown in \cref{fig:OptimalAlphaBeta}(b). 
Here, overdamping occurs for $\beta \gtrsim 20$. 
If $\alpha$ or $\beta$ is chosen too large, the time integration becomes numerically unstable, and the \respch{cooling} dynamics can no longer be reliably followed. 
These results indicate that intermediate values of $\alpha$ and $\beta$ provide the most efficient cooling dynamics without compromising stability.

\section{\respch{Cooling} terms for the generalized single-particle Hamiltonian}\label{appendix-C}

In the main text, we derived the \respch{cooling} terms under the simplifying assumption that the single-particle Hamiltonian takes the basic form \cref{eq:hss}.
Within a density functional theory framework, however, the single-particle Hamiltonian typically acquires a more general structure, namely
\begin{equation}\label{eq:hssgen}
  h_{\sigma\sigma} = -\vect{\nabla}\frac{1}{2m^*_\sigma(\vect{r})}\vect{\nabla} + U_{\sigma}(\vect{r}) -\frac{i}{2}\{\vect{A}_\sigma(\vect{r}),\vect{\nabla}\},
\end{equation}
where $\{\cdot,\cdot\}$ denotes the anticommutator.
Here $m^*_\sigma(\vect{r})$ is a possibly position-dependent effective mass, and $\vect{A}_\sigma(\vect{r})$ is a vector potential (e.g., associated with a magnetic field) that supplements the scalar potential $U_\sigma(\vect{r})$.

With this form, the expression \cref{eq:comm_hrho} for \respch{cooling} mean-field potential generalizes to (suppressing explicit position dependence for clarity):
\begin{align}\label{eq:comm_hrho_gen}
    \Udiss_{\sigma}=&-i \tilde{\alpha}\,\diag\Bigl(\big[ h_{\sigma\sigma}, \rho_{\sigma\sigma} \big]\Bigr)\nonumber\\
    =&-\frac{\tilde{\alpha}}{m^*_\sigma}\vect{\nabla}\cdot\vect{j}_\sigma -\tilde{\alpha}\vect{\nabla}\left(\frac{1}{m^*_\sigma}\right)\cdot\vect{j}_\sigma\nonumber\\
    &-2\tilde{\alpha}\vect{A}_\sigma\cdot\sum_n\text{Re}\left[v_{n,\sigma}^*\vect{\nabla}v_{n,\sigma}\right]-\tilde{\alpha}\rho_{\sigma}\vect{\nabla}\cdot\vect{A}_\sigma,
\end{align}
where $\rho_{\sigma} \equiv \rho_{\sigma\sigma}(\vect{r}, \vect{r})$ is the local density.
It correctly reduces to \cref{eq:comm_hrho} in the special case of a constant effective mass, $m^*(\vect{r})=m$, and vanishing vector potential, $\vect{A}_\sigma=0$, when setting $\tilde{\alpha}=\alpha/\rho_0$. 

Similarly, the expression \cref{eq:diagB} generalizes to
\begin{align}
    B' =& -\frac{1}{2m^*_{\ua}}\sum_n u_{n\da}\nabla^2v_{n\ua}^* -\frac{1}{2m^*_{\da}}\sum_n v_{n\ua}^*\nabla^2u_{n\da} \nonumber \\
    &-\left[\frac{1}{2}\vect{\nabla}\left(\frac{1}{m^*_\ua}\right)+\I\vect{A}_\ua\right]\cdot\sum_n u_{n\da}\vect{\nabla} v_{n\ua}^*\nonumber\\
    &-\left[\frac{1}{2}\vect{\nabla}\left(\frac{1}{m^*_\da}\right)+\I\vect{A}_\da\right]\cdot\sum_n v^*_{n\ua}\vect{\nabla} u_{n\da}\nonumber\\
    & -\frac{i}{2}\vect{\nabla}\cdot\left(\vect{A}_\ua+\vect{A}_\da\right)\kappa,
\end{align}
where $\kappa=\kappa_{\ua\da}(\vect{r},\vect{r})$.
This expression reduces to \cref{eq:diagB} in the limit of constant effective mass and vanishing vector potential.
The formulas \cref{eq:Deltadiss1,eq:Deltadiss2} remain unchanged, except that $\theta_{\mathscr{B}^{\prime}}$ must now be computed using the above generalized form.

% \bibliography{macros,main}% Produces the bibliography via BibTeX.

\end{document}